\documentclass[acmlarge]{acmart}
\AtBeginDocument{%
  }

\setcopyright{acmlicensed}
\copyrightyear{2025}
\acmYear{2025}
\acmDOI{XXXXXXX.XXXXXXX}

\acmJournal{IMWUT}
\acmVolume{0}
\acmNumber{0}
\acmArticle{0}
\acmMonth{8}

\usepackage[caption=false,font=normalsize,labelfont=sf,textfont=sf]{subfig}
\usepackage{amsfonts}
\usepackage{makecell}
\usepackage{multirow}
\usepackage{multicol}
\usepackage{xspace}
\usepackage{amsmath}
\newcommand{\name}{\textsc{GesFi}\xspace}
\usepackage{tcolorbox}

\begin{document}

\title{Beyond Physical Labels: Redefining Domains for Robust WiFi-based Gesture Recognition}


\author{Xiang Zhang}
\email{zhangxiang@ieee.org}
\orcid{0000-0003-0413-6135}
\affiliation{%
  \institution{Tianjin University}
  \city{Tianjin}
  \country{China}}

\author{Huan Yan}
\authornote{Corresponding authors}
\email{yh1995.cs@gmail.com}
\orcid{0009-0008-1810-4920}
\affiliation{%
  \institution{Guizhou Normal University}
  \city{Guiyang}
  \country{China}}

\author{Jinyang Huang}
\authornotemark[1]
\email{hjy@hfut.edu.cn}
\orcid{0000-0001-5483-2812}
\affiliation{%
  \institution{Hefei University of Technology}
  \city{Hefei}
  \country{China}}  

\author{Bin Liu}
\email{flowice@ustc.edu.cn}
\orcid{0000-0002-3977-8800}
\affiliation{%
  \institution{University of Science and Technology of China}
  \city{Hefei}
  \country{China}}  

\author{Yuanhao Feng}
\email{fy-hace@mail.ustc.edu.cn}
\orcid{0000-0003-3969-9472}
\affiliation{%
  \institution{The University of Electro-Communications}
  \city{Tokyo}
  \country{Japan}} 

\author{Jianchun Liu}
\email{jcliu17@ustc.edu.cn}
\orcid{0000-0002-1764-9303}
\affiliation{%
  \institution{University of Science and Technology of China}
  \city{Hefei}
  \country{China}}  

\author{Meng Li}
\email{mengli@hfut.edu.cn}
\orcid{0000-0003-3553-0813}
\affiliation{%
  \institution{Hefei University of Technology}
  \city{Hefei}
  \country{China}}  

\author{Fusang Zhang}
\email{zhangfusang@buaa.edu.cn}
\orcid{0000-0002-2529-8021}
\affiliation{%
  \institution{Institute of Software, Chinese Academy of Sciences}
  \city{Beijing}
  \country{China}
}
\affiliation{%
  \institution{Inspur Computer Technology Co., Ltd}
  \city{Beijing}
  \country{China}
}

\author{Zhi Liu}
\email{liu@ieee.org}
\orcid{0000-0003-0537-4522}
\affiliation{%
  \institution{The University of Electro-Communications}
  \city{Tokyo}
  \country{Japan}} 
\renewcommand{\shortauthors}{Zhang et al.}

\begin{abstract}

WiFi-based gesture recognition holds great promise due to its non-intrusive and ubiquitous nature. However, robust generalization across diverse conditions remains a core challenge due to the inherent sensitivity and ambiguity of Channel State Information (CSI) signals, making them unreliable for real-world human-computer interaction (HCI) applications. While domain adversarial learning-based systems have proven effective in other sensing modalities, its performance in WiFi-based gesture recognition remains limited.
In this paper, we revisit the foundational assumptions in current adversarial learning-based WiFi gesture recognition systems and identify a key mismatch: conventional methods rely on subjective physical domain labels (e.g., location, orientation), which often fail to reflect true distributional patterns in the data. This misalignment introduces two critical issues, namely classification conflict and manifold distortion, which fundamentally limit generalization performance.
To address these challenges, we propose~\name, a novel WiFi-based gesture recognition system that introduces WiFi latent domain mining to redefine domains directly from the data itself. \name first processes raw sensing data collected from WiFi receivers using CSI-ratio denoising, Short-Time Fast Fourier Transform, and visualization techniques to generate standardized input representations. It then employs class-wise adversarial learning to suppress gesture semantic and leverages unsupervised clustering to automatically uncover latent domain factors responsible for distributional shifts. These latent domains are then aligned through adversarial learning to support robust cross-domain generalization. Finally, the system is applied to the target environment for robust gesture inference.
We deployed~\name under both single-pair and multi-pair settings using commodity WiFi transceivers, and evaluated it across multiple public datasets and real-world environments. Compared to state-of-the-art baselines, \name achieves up to 78\% and 50\% performance improvements over existing adversarial methods, and consistently outperforms prior generalization approaches across most cross-domain tasks. (Code available at \url{https://github.com/CamLoPA/GesFiCode}).

\end{abstract}



\received{20 February 2007}
\received[revised]{12 March 2009}
\received[accepted]{5 June 2009}

\maketitle

\section{Introduction}


Gesture recognition systems~\cite{huang2024keystrokesniffer} play a pivotal role in the field of human-computer interaction. In recent years, WiFi-based gesture recognition systems have garnered significant attention due to its universal deployment capabilities and non-intrusive nature~\cite{gao2023wicgesture,chen2024wignn}, and the integration of machine learning has further enhanced the performance of these systems. However, WiFi sensing frequently faces major generalization challenges~\cite{liu2024unifi,sheng2024metaformer}. 
At the core of WiFi-based gesture recognition is the sensing of limb-induced changes in signal reflection paths, captured through Channel State Information (CSI)~\cite{li2025wilife,yi2024enabling,meng2023secur}. This mechanism is intrinsically low-dimensional, as it compresses the three-dimensional spatial dynamics of hand movements into a one-dimensional representation of reflection path length.
Consequently, WiFi sensing exhibits two critical characteristics: high sensitivity to various factors (e.g., location, orientation) and significant semantic ambiguity. A single gesture may produce dramatically different CSI patterns under varying conditions, while distinct gestures may yield similar CSI traces. These properties make domain shifts in WiFi sensing particularly complex.

\begin{figure}[tp]
\centering
\includegraphics[width=0.9\linewidth]{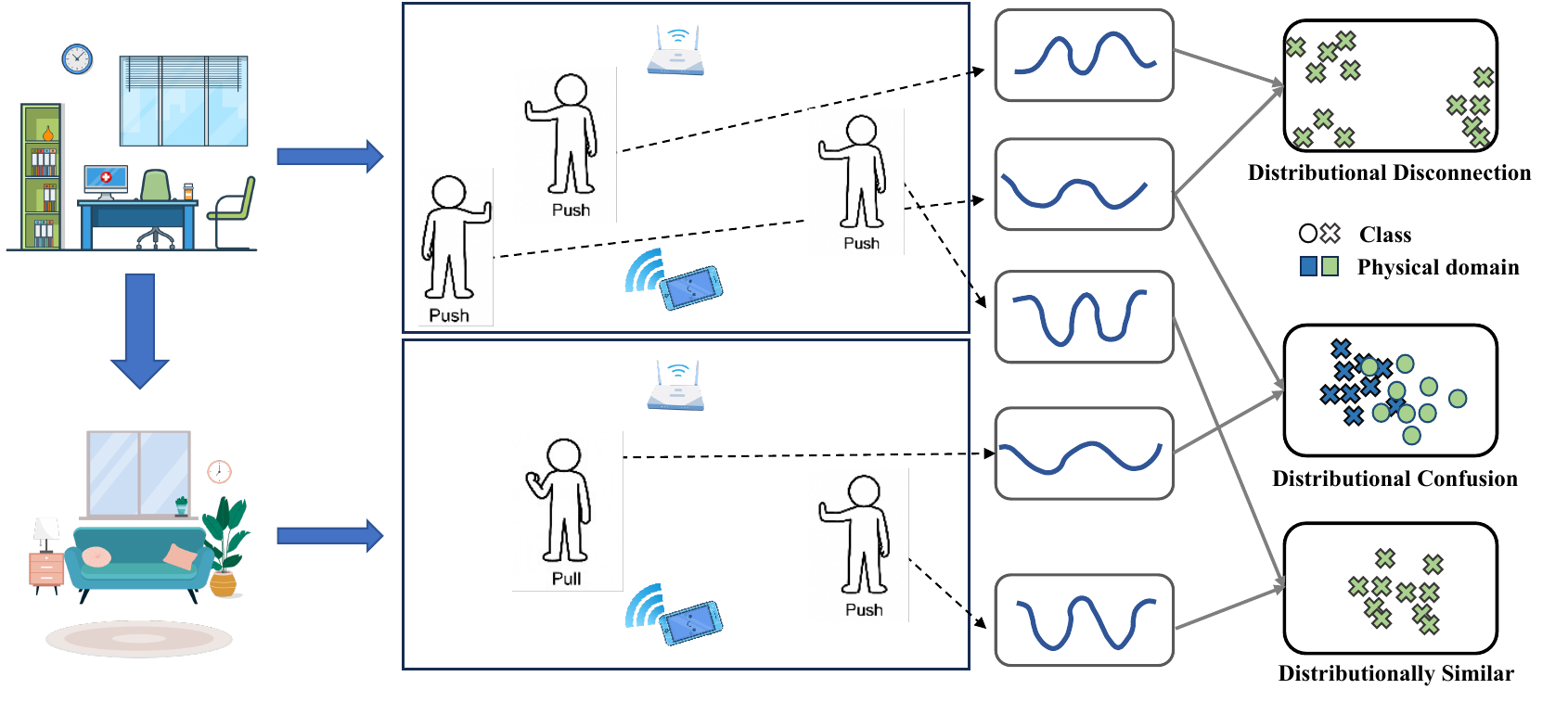}
\caption{Prior approaches~\cite{jiang2018towards} typically assume that distributional shifts can be captured by physical labels, such as environments. However, we argue that physical domains are insufficient to represent cross-domain distributional variations in WiFi sensing. For example, within a single environment, a push gesture performed in opposite directions can result in entirely different CSI patterns, forming multiple disconnected distributional clusters. Conversely, push and pull gestures executed in different environments may exhibit highly similar CSI representations, leading to distributional confusion. In some cases, the variation in CSI within a single physical domain may exceed that between different domains.} 
\label{fig:intro}
\vspace{-0.15in}
\end{figure}

A straightforward approach to enhancing generalization would involve collecting large and diverse WiFi sensing datasets that capture a wide range of potential environmental and user conditions. However, collecting and annotating wireless sensing data is highly cumbersome compared to visual or auditory data, severely limiting the scale and diversity achievable in practice. Consequently, recent research has increasingly focused on developing generalized gesture recognition systems using limited training data.~\cite{sheng2024cdfi}. Existing strategies fall into three broad categories: domain-invariant feature extraction, domain adaptation, and domain generalization. Domain-invariant feature extraction leverages expert knowledge to manually design features less sensitive to domain shifts, such as the Body-coordinate Velocity Profile (BVP)~\cite{zhang2021widar3} and the Motion Navigation Primitive (MNP)~\cite{gao2021towards}. However, such handcrafted features often lead to non-negligible information loss compared to the original signal. Domain adaptation~\cite{zhou2022domain,zhang2023unsupervised} enables gesture recognition systems to adjust to target domains via distribution alignment, assuming access to target domain data. Yet, even when only a few samples per class are required~\cite{sheng2024metaformer,feng2022wi}, the overall data collection burden remains substantial in complex WiFi environments.

In contrast, domain generalization aims to build gesture recognition systems that can generalize effectively to unseen domains using limited training data, by leveraging the powerful inductive capabilities of machine learning.
Existing approaches have primarily pursued this goal by designing more expressive architectures capable of extracting richer signal representations~\cite{liu2024unifi, gu2022wigrunt}, or by augmenting the source domain distribution through synthetic or transformed data~\cite{hou2024rfboost,chi2024rf}.
Some methods have also adopted domain adversarial learning as a means to achieve domain generalization in WiFi-based gesture recognition~\cite{jiang2018towards, liu2023wisr, hassan2024adversarial}. These approaches typically rely on domain labels to distinguish source domains and employ a minimax game between a feature extractor and a domain discriminator to learn domain-invariant representations.
Due to its intuitive mechanism for eliminating domain-specific factors, along with its theoretical guarantees and framework flexibility, domain adversarial learning has become a widely used strategy across various sensing modalities.
However, multiple recent studies~\cite{liu2024unifi, zhang2025wiopen, liu2023generalizing, yan2025wisfdar} have reported suboptimal performance when applying adversarial-based methods to WiFi gesture recognition~\cite{jiang2018towards, liu2023wisr, hassan2024adversarial}.
This naturally leads to the following question:

\begin{tcolorbox}[colback=gray!25!white, size=title, boxsep=1mm, colframe=white, before={\vskip1mm}, after={\vskip0mm}]
\textbf{Q1: Why do adversarial-based WiFi gesture recognition systems exhibit limited effectiveness?}
\end{tcolorbox}


We posit that the limited effectiveness of existing approaches stems from a fundamental misconception regarding the notion of \textbf{"domain"} in WiFi sensing. Traditionally, WiFi gesture recognition systems aim to generalize across unseen physical domains—such as new environments, locations, or user orientations, and accordingly, most domain-adversarial methods have been designed based on these physical labels.
However, we argue that in the context of WiFi sensing, physical domains do not align with the type of domain differentiation that adversarial learning is designed to address. Specifically, domain adversarial learning seeks to align data distributions in the feature space, not domains that differ primarily in their physical semantics.
This technique was originally developed for high-dimensional modalities like vision, where domain labels (e.g., real vs. cartoon images) tend to correspond closely with data distributional shifts. 


In contrast, WiFi sensing exhibits fundamentally different characteristics, its low-dimensional, and noisy nature often causes physical labels to poorly reflect the true distributional structure of sensing data.
As illustrated in Figure~\ref{fig:intro}, even within the same physical environment, variations in user position, orientation, and unmodeled multipath effects can introduce greater variability in CSI than the environment change itself. As a result, samples within a single physical domain frequently form multiple scattered and disconnected clusters in the data space.
At the same time, semantically distinct gestures performed across different physical domains may produce highly similar CSI patterns, a phenomenon we refer to as semantic conflict.
Moreover, due to the inherent dimensionality reduction in WiFi sensing, it becomes extremely difficult to infer meaningful physical domain shift trajectories directly from CSI data, especially when domains differ significantly in their spatial configurations.
In contrast, such transitions are often more interpretable in vision tasks (e.g., transforming real images into cartoons), where domain shifts unfold along continuous and semantically coherent paths.
Consequently, adversarial-based approaches in WiFi sensing face the following fundamental limitations (Theoretical analysis see Section~\ref{sec:adv}):

\textbf{Classification Conflicts:} 
Deep learning models are inherently prone to shortcut learning—they tend to minimize loss using the most accessible cues in the data, even if those cues are semantically spurious. In the presence of semantic conflicts, domain adversarial learning tends to amplify the confusion between these gestures, as increasing their indistinguishability makes it easier to minimize adversarial loss.
However, effective gesture recognition requires maximizing inter-class separability to ensure accurate classification. This leads to a fundamental conflict between the goals of adversarial domain alignment and gesture discrimination.

\textbf{Manifold Distortion:} 
Domain adversarial learning enforces distributional alignment across every source domains. However, when two domains differ substantially in their physical sensing perspectives (e.g. user orientation), it becomes extremely difficult to capture the true domain shift trajectory using only one-dimensional CSI signals.
In such cases, deep learning models tend to favor alignment paths that are easier to optimize statistically, rather than those that reflect meaningful physical transitions. This results in distorted domain manifolds that fail to preserve the geometric structure of actual physical changes.

Motivated by these insights, we advocate for a fundamental rethinking of domain definition in WiFi-based recognition systems. Our key observation is that domain generalization aims to eliminate the factors that cause distributional shifts. In WiFi sensing, however, such shifts often stem from complex and entangled combinations of physical factors, making it difficult to delineate domains using simple physical labels alone.
This raises a natural question:\textbf{ why not identify domain differences (latent domains) directly from the distributional structure of the sensing data itself?}
By doing so, we can intentionally assign semantically conflicting gestures to the same latent domain to avoid optimization conflicts, and interpolate between physically dissimilar subdomains through intermediate latent domains—thus preventing the direct alignment of highly divergent physical factors. This strategy effectively mitigates the aforementioned limitations of semantic conflict and manifold distortion.


To this end, we propose~\name, a WiFi-based gesture recognition system that leverages WiFi latent domain mining for robust domain generalization. \name first captures gesture-relevant CSI from commodity WiFi transceivers. Then, we apply the CSI-ratio method~\cite{zeng2019farsense} for denoising, which produces cleaner amplitude and phase signals. Subsequently, the Short-Time Fourier Transform (STFT) is applied to extract Doppler Frequency Shifts (DFS). Following the procedure of UniFi~\cite{liu2024unifi}, the phase and DFS obtained from each antenna pair are visualized, concatenated, and resized to construct a unified, high-quality input representation.
In the generalization learning phase, \name initially employs class-label-based adversarial training to suppress gesture semantics. It then performs unsupervised clustering to automatically uncover latent domain factors that drive distributional shifts, and partitions the data accordingly. These latent domains are subsequently aligned through adversarial domain generalization. Finally, gesture recognition is performed using CSI collected in the target environment.


We deploy \name under both single-pair and multi-pair configurations using commodity WiFi devices, and evaluate it on three public datasets and in real-world environments. Compared to domain-adversarial baselines such as EI~\cite{jiang2018towards} and WiSR~\cite{liu2023wisr}, \name achieves performance improvements of over 78\% and 50\%, respectively. It also consistently outperforms other state-of-the-art methods across most cross-domain tasks.

Our key contributions are summarized as follows:

\begin{itemize}
  \item We are the first to systematically reveal, both theoretically and empirically, the root causes behind the limited effectiveness of adversarial generalization module in current WiFi gesture recognition systems. We challenge the prevailing understanding of “domain” in this field and identify two critical limitations, namely semantic conflict and manifold distortion, that arise from the traditional physical domain definition.
  

  \item We propose \name, a novel WiFi-based gesture recognition system based on a new perspective: objectively mining "latent domains" directly from the data's distribution. Our approach automatically discovers the factors causing distribution shifts, and avoiding biases introduced by physical labels. \name effectively harnesses the power of adversarial learning, significantly boosting the system's generalization capabilities.
  
  \item We implement \name on commodity WiFi hardware in both single-pair and multi-pair configurations, and evaluate it on three public datasets and a real-world environment. Results show substantial improvements over both domain-adversarial and state-of-the-art baselines across almost all cross-domain tasks.
\end{itemize}

\section{Preliminary}

\subsection{WiFi CSI Based Gesture Recognition}
Current research in WiFi sensing primarily relies on CSI~\cite{zhang2023wital,li2024uwb,wang2025vr}. CSI describes the effects such as attenuation experienced by a signal as it propagates from the transmitter to the receiver~\cite{yi2025multi,zhang2025camlopa,zhang2025diffloc}. The propagation process of WiFi signals can be represented as follows:
\begin{equation}
	\label{equ:CH}
	R= H S+ \mathcal{N}
	\end{equation}
Where $R$ and $S$ are the received and transmitted signal vectors, respectively. $\mathcal{N}$ is additive white Gaussian noise and $H$ is the CSI channel matrix.

At the receiver, the CSI obtained is a combination of multiple propagation paths, which can be divided into dynamic and static components. Static components refer to signals propagated through the Line-of-Sight (LoS) path and signals reflected by static objects such as furniture and walls. On the other hand, dynamic components are signals reflected by moving objects, such as a person's arms or legs, while performing gestures.
\begin{equation}
\label{equ:CFRSUM}
H(f,t)=H_s(f,t)+H_d(f,t)
\end{equation}
Where $f$ and $t$ represent the signal frequency and the timestamp, respectively. The dynamic CSI can be further elaborated as follows~\cite{huang2021phaseanti,xu2024hypertracking,tong2024nne}:
\begin{equation}
\label{equ:CFR}
H_d(r,t)=\sum_{k\in \mathbf{D}} h_k(r,t) e^{-j2\pi \frac{ d_k(t)} {\lambda_k}}
\end{equation}
Here, $h_k(r,t)$, $d_k(t)$, and $\lambda_k$ represent the attenuation, the path length of the dynamic path, and the wavelength associated with the $k^{th}$ path, respectively. The set $\mathbf{D}$ encompasses dynamic paths. In WiFi-based gesture recognition, CSI implicitly encodes gesture dynamics (movement of hands and arms) in the path length of dynamic components. Such encoding is inherently low-dimensional: it primarily captures changes in path lengths, rather than full 3D spatial information. Consequently, WiFi sensing exhibits two critical properties: \textbf{High sensitivity.} Minor environmental or body movement variations can cause large signal changes. \textbf{Significant ambiguity.} Different gestures may generate similar CSI, and the same gesture may yield different CSI.


\subsection{Problem Definition}

We first introduce some notation used in this paper. Let $\mathcal{X}$ denote the input space and $\mathcal{Y}$ represent the label space. A domain can be defined as the joint distribution $P(X,Y)$ of $\mathcal{X}$ and $\mathcal{Y}$. For a domain $P(X,Y)$, we consider $P_{X}$ as the marginal distribution of $X$, and $P_{Y|X}$ as the posterior probability of $Y$ given $X$.

\textbf{In-Domain Gesture Recognition.} Many previous studies have focused on in-domain WiFi-based gesture recognition, where only one domain (denoted as $S = \left \{x,y \right \}$) exists. Both the training and testing data are drawn from the same underlying distribution $P(X,Y)$, meaning that the sensing environment and configuration used for data collection are identical during both training and evaluation. The objective of this task is to train a classification model on the training data.
\begin{equation}
\label{equ:INF}
\hat{y} =f(x)
\end{equation}
where $x$ represents input samples, and $\hat{y}$ represents the predicted labels. Existing deep learning-based methods typically encourage the model to evolve towards fitting the distribution of $P(X,Y)$ through the loss function $L(f(x),y)$ and the ground truth label $y$.

\textbf{Domain Generalization Gesture Recognition.} In domain generalization scenarios, we have access to K similar but different source domains (denoted as $\mathcal{S}=\left \{\mathcal{S}_k = \left \{x^k,y^k \right \}\right \}$), and each source domain is associated with a joint distribution $P^k(X,Y)$. It is worth mentioning that $P^i(X,Y) \ne P^j(X,Y)$ with $i \ne j$ and $i,j \in \left \{ 1,...,K\right\}$. For our task, this means that these domains share identical gesture classes and are governed by the same sensing principles, which ensures comparable feature spaces. However, their data distributions differ due to multipath propagation and spatial variations. The goal of domain generalization gesture recognition is to learn a model $\hat{y} =f(x)$ using only data in $\mathcal{S}$ which satisfies:
\begin{equation}
\label{equ:INF1}
\min_{x^t\in \mathcal{T}}L(f(x^t),y^t) 
\end{equation}
Where $\mathcal{T}$ means unseen target domain and the corresponding joint distribution is $P^\mathcal{T}(X,Y)$. In particular, the target domain also satisfies $P^\mathcal{T}(X,Y) \ne P^k(X,Y)$ for $k \in \left \{ 1,...,K\right\}$, meaning that its data distribution differs from those of the source domains. For instance, the target domain may include previously unseen orientations, environments, or locations. However, it is important to note that the target domain remains similar to the source domains, sharing the same gesture classes and sensing principles. When the gap between domains becomes excessively large and no auxiliary knowledge is available, domain generalization methods are no longer effective.

\section{Domain Adversarial-based Generalization in WiFi Sensing}
\label{sec:adv}

\subsection{Preliminary Experiments}
\begin{figure}
  \centering
  \includegraphics[width=0.48\linewidth]{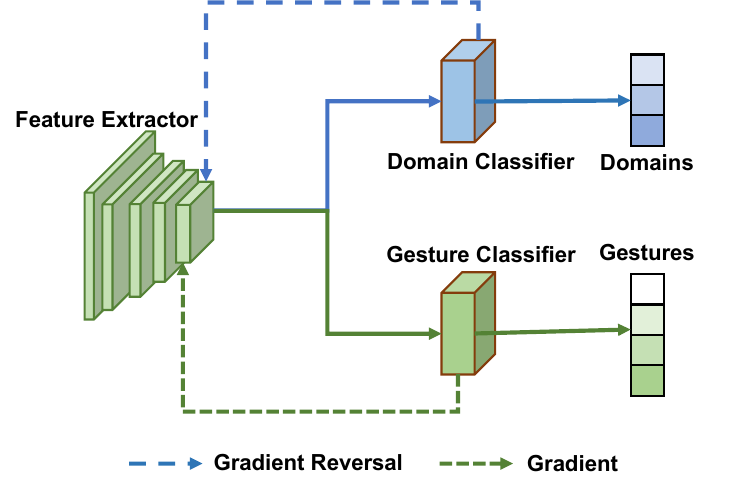}
  \caption{The domain adversarial learning based generalization framework.}
  \label{fig:grl}
\end{figure}

Currently, adversarial-based methods have become the main approach to domain generalization due to their solid theoretical foundations and flexibility in application. As depicted in Figure \ref{fig:grl}, this framework~\cite{ganin2015unsupervised} employs a feature extractor to extract features from the data, a classifier to identify gesture categories, and a domain discriminator to discern the domain label of the data. Through the gradient reversal operation, this framework can eliminate the information deemed effective by the domain discriminator, which can effectively encourage the feature extractor to extract domain-agnostic but gesture-related features. However, compared to the main solutions in WiFi gesture recognition~\cite{gu2022wigrunt,zhang2025wiopen,liu2024unifi,sheng2024metaformer,li2020wihf}, adversarial-based methods~\cite{jiang2018towards,liu2023wisr} have not achieved competitive performance, which we attribute to the use of coarse-grained physical domain labels in these methods.

\begin{figure}[htp]
	\centering
		\subfloat[Cross-Orientation]{\label{ro}
		\begin{minipage}{0.46\linewidth}
			\centering
			\includegraphics[width=1\textwidth]{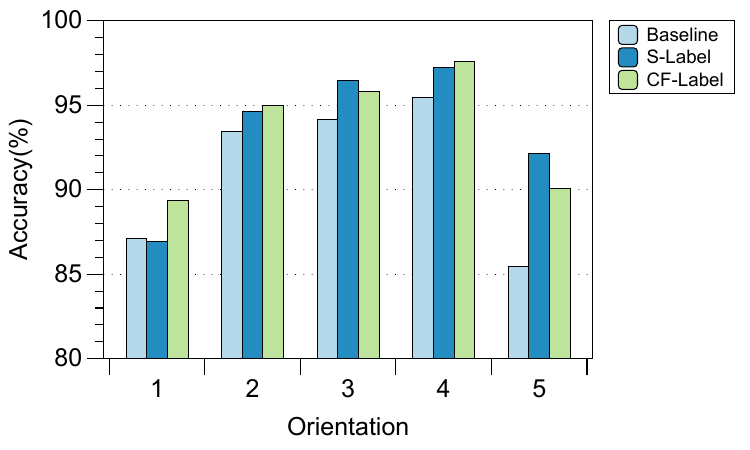}
		\end{minipage}
	}	
	\quad
	\subfloat[Cross-Location]{\label{rl}
		\begin{minipage}{0.46\linewidth}
			\centering
			\includegraphics[width=1\textwidth]{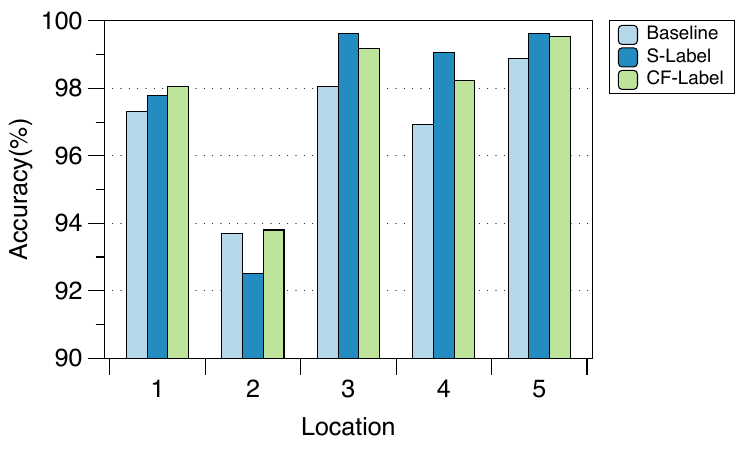}
		\end{minipage}
	}	
	\quad
	\caption{Exploratory experiments results. S-label and CF-label represent cross domain with physical domain labels and counter-intuitive physical domain labels, respectively.}
	\label{fig:grlanddg}
	\vspace{-0.0in}
\end{figure}

To empirically motivate our theoretical analysis, we conducted preliminary experiments to investigate the relationship between physical domain partitioning and adversarial generalization performance in WiFi sensing. We conducted experiments on the Widar3.0 dataset using the aforementioned framework (Figure~\ref{fig:grl}, which is also used by~\cite{jiang2018towards,liu2023wisr}). The experiments employ a ResNet18 backbone pre-trained on ImageNet, using visualized CSI images as input. The data visualization method and the backbone architecture are consistent with various existing works \cite{gu2022wigrunt,liu2024unifi}. We conducted separate experiments for cross-orientation and cross-location tasks. In the Widar3.0 dataset, each sample is labeled with location and orientation, and comprises data collected from $5$ locations/orientations. Our experiments follow the leave-one-out principle, where $4$ sets of location/orientation data are used for training, and the remaining $1$ is used for testing. The descriptions of the three settings adopted in the experiments are as follows:
\begin{itemize}
    \item \textbf{Baseline.} The baseline gesture classifier is constructed using ResNet18 and a fully connected layer.
    \item \textbf{Cross domain with physical domain labels.} The adopted network architecture is illustrated in Figure \ref{fig:grl}. The physical discrepancy between the testing and training data domains is utilized as domain labels, such as using orientations as domain labels in the cross-orientation task.
    \item \textbf{Cross domain with counter-intuitive physical domain labels.} The network architecture employed is depicted in Figure \ref{fig:grl}. Domain labels are assigned based on information contrary to physical intuition, such as using location labels as domain labels in the cross-orientation task.
\end{itemize}

The experiment results are shown in Figure \ref{fig:grlanddg}, where we observe three kinds of results:

\begin{itemize}
    \item \textbf{Physical label performance superiority:} 
    The physical label achieves the best performance, particularly dominating in relatively straightforward cross-location task (Location 3,4,5). This observation aligns well with our cognitive, indicating the effectiveness of leveraging physical domain discrepancy between the target and source domains through adversarial gradients to enhance generalization performance.
    \item \textbf{Physical label outperforms the baseline but is inferior to the counter-intuitive physical label:} While the performance of the physical label surpasses the baseline, it is inferior to the counter-intuition physical label, particularly prevalent in more challenging cross-orientation experiments (Orientation 2,4). This finding contradicts intuition, as it suggests that features extracted from seemingly unrelated domain adversarial mechanisms outperform those derived from physical domains.
    \item \textbf{Baseline outperforms physical label:} In both cross-location and cross-orientation experiments, some instances are observed where the baseline outperforms the physical label in the most challenging tasks (Orientation 1 and Location 2). This outcome, again contrary to intuition, implies that the physical domain labels even induce negative outcomes.
\end{itemize}

The above results suggest that domain labels based on physical factors may not yield the expected generalization performance in WiFi gesture recognition tasks, and may even introduce adverse effects. In the next section, we theoretically explore the reasons for this and present our proposed solution.

\subsection{How to Make Adversarial-based Generalization More Effective in WiFi Sensing?}

We first provide some preliminaries to explain the fundamental principles of adversarial learning–based WiFi sensing frameworks, analyze their inherent limitations, and present the motivation behind our proposed approach.

\textbf{Theorem 1} (Theorem 2.1 in \cite{sicilia2023domain}, modified from Theorem 2 in \cite{ben2010theory}). Let $\mathcal{X}$ be a space, and $\mathcal{F}$ be a class of hypotheses. Suppose $S$ and $\mathcal{T}$ are distributions of the source and target domain over $\mathcal{X}$. Then for any $f\in \mathcal{F}$, the following holds:
\begin{equation}
     \varepsilon_\mathcal{T} (f) \le \lambda +  \varepsilon_\mathcal{S} (f) + \frac{1}{2} d_\mathcal{H}(\mathcal{S},\mathcal{T})
\end{equation}
The $\varepsilon (f)$ is the error and $d_\mathcal{H}()$ is the $\mathcal{H}$-divergence, for more explanation, please refer to Appendix~\ref{app}. This theorem provides an upper bound on the generalization error in the target domain, where $\lambda$ is typically negligible. It highlights that achieving accurate target-domain WiFi gesture recognition requires both minimizing the source-domain error and reducing the domain divergence $d_\mathcal{H}(\mathcal{S},\mathcal{T})$, which underscores the importance of learning domain-invariant features.

Based on Theorem 1, the following proposition can be obtained.

\textbf{Proposition 1} (Proposition 3.2 in \cite{lu2024diversify}). Let $\mathcal{X}$ be a space and $\mathcal{F}$ be a class of hypotheses. Suppose $S=\left \{S_k = \left \{x^k,y^k \right \}\right \}$ and $\mathcal{T}$ are distributions of the source and target domain over $\mathcal{X}$. Then for any $f\in \mathcal{F}$, the following holds:
\begin{equation}
\label{eq:9}
     \varepsilon_\mathcal{T} (f) \le \lambda +  \sum_{k}^{}\varphi _k \varepsilon_{\mathcal{S}^k} (f) + \frac{1}{2} d_\mathcal{H}(\mathcal{S},\mathcal{T}) + \frac{1}{2}\max_{i,j} d_\mathcal{H}(\mathcal{S}^i,\mathcal{S}^j)
\end{equation}
Proposition 1 extends Theorem 1 by incorporating multiple subdomains within the source data and emphasizes that reducing divergence both among source subdomains and between source and target domains can further tighten the generalization bound. In WiFi-based sensing, these subdomains correspond to CSI data collected under varying physical conditions such as different users, locations, or orientations. Existing adversarial domain generalization methods in WiFi gesture recognition are designed to achieve this objective by encouraging a shared feature space in which samples from different physical domains become indistinguishable to a domain discriminator. In practice, this process implicitly minimizes the maximum divergence among source subdomains (that is, $\frac{1}{2}\max_{i,j} d_\mathcal{H}(\mathcal{S}^i,\mathcal{S}^j)$), which consequently tightens the generalization error bound.

It is important to note that the conditions in Theorem 1 and Proposition 1 assume that the divergence between any pair of domains (source–source or source–target) is bounded by a finite constant, ensuring that their latent representations share partially overlapping feature supports. In the context of WiFi sensing, this means that all domains are collected under comparable sensing configurations (e.g., same sensing principles, and gesture semantics), although location or orientation differences cause distributional shifts in CSI. These bounded variations define what we refer to as generalizable domains.



In the following, we investigate two key challenges arising from conventional physical domain partitioning in WiFi-based gesture recognition, and explain how they fundamentally limit adversarial generalization: (1) classification conflicts caused by signal ambiguity, and (2) manifold distortion caused by complex domain transition paths. Let $\mathcal{X_c}$ denote the WiFi sensing distribution space and $\mathcal{Y_c}$ the gesture label space. Samples $(x_c, y_c)$ are drawn from a joint distribution $\mathcal{D_c}$ over $\mathcal{X_c} \times \mathcal{Y_c}$. Each sample $x_c$ is generated as:
\begin{equation}
    x_c=s(y_c,d)+\epsilon
\end{equation}
where $d$ is a physical domain factor (e.g., location, orientation, environment), $s$ models the sensing mechanism, and $\epsilon$ represents noise and other stochastic perturbations.

Due to the inherently ambiguous nature of WiFi sensing signals, samples from different physical domains that correspond to different gesture classes $y_c$ may easily overlap in the feature space. We define $\eta(\mathcal{S}^i, \mathcal{S}^j)$ (see Appendix~\ref{app} for details) as the probability that two samples drawn from different domains but close in the feature space belong to different gesture classes. During adversarial training with gradient reversal, the domain discriminator promotes similarity among samples from different domains to confuse the domain classifier, while the task classifier simultaneously attempts to separate samples of distinct classes. Consequently, in regions where $\eta$ is high, adversarial alignment and classification objectives become conflicting, leading to an increase in the source-domain error $\varepsilon_S(h)$. Then, we have the following proposition.

\textbf{Proposition 2 (Classification Conflict):} If $\eta(\mathcal{S}^i, \mathcal{S}^j)$ is high, adversarial alignment in feature space increases $\sum_{k}^{}\varphi _k \varepsilon_{\mathcal{S}^k} (f)$ due to optimization conflict. (More details please refer to Appendix~\ref{app}.)


WiFi physical domain factors change in highly nonlinear, intricate trajectories. Physical domains with significant differences (e.g., opposite gesture directions) may lie on very distant regions of the true underlying data manifold. Naïve adversarial alignment aligns all source subdomains indiscriminately, forcing feature matchings between physically incompatible distributions, then the following proposition can be obtained.

\textbf{Proposition 3 (Manifold Distortion):} If two physical domains $\mathcal{S}^i, \mathcal{S}^j$ are separated by a complex manifold trajectory, adversarial alignment directly between $\mathcal{S}^i$ and $\mathcal{S}^j$ introduces feature space distortion, causing $d_\mathcal{H}(\mathcal{S}_i, \mathcal{T})$ to be misestimated. (More details please refer to Appendix~\ref{app}.)

The preceding analysis demonstrates that conventional physical domain partitioning is theoretically inadequate for achieving effective adversarial generalization in WiFi-based gesture recognition. While it is conceivable to mitigate these issues through extremely fine-grained physical labeling, such an approach is often impractical due to prohibitive annotation costs, increased computational complexity, and the fundamental difficulty of accurately modeling intricate physical influences. To address these challenges, we propose a simple yet effective solution: \textbf{WiFi Latent Domain Mining}. Our key insight is that the distribution of WiFi sensing data is shaped jointly by class identity and domain-specific characteristics, but the true domain boundaries are better revealed through the data distribution itself rather than imposed by subjective physical factors. The ultimate goal of domain generalization is to learn a unified feature space that remains robust against domain shifts. Therefore, rather than relying on physical-perspective partitioning, we adopt a data-driven strategy grounded in the worst-case principle: partitioning the overall sensing data into several distinct latent domains directly based on distributional structures. This approach enables us to automatically and effectively summarize the key domain factors responsible for domain distribution shifts, thereby enhancing adversarial generalization by satisfying the following properties:

\begin{itemize}
    \item \textbf{Reducing classification conflicts:} By grouping data with different categories but similar distributions into a single latent domain, minimizing $\eta$. Which ensuring that gesture samples with similar distributional characteristics are assigned to the same latent domain, thereby effectively prevents the adversarial module and the gesture classifier from conflicting when handling such samples.
    \item \textbf{Mitigating manifold distortion:} The latent domain mining integrates disparate physical domains into intermediate-state domains, thereby avoiding the direct alignment of domains that differ excessively.
\end{itemize}

In implementing our approach, the number of latent domains must strike a balance between modeling capacity and fitting complexity. If too few latent domains are used, the model may fail to capture the diversity and richness of the underlying data distribution. Increasing the number of latent domains allows the model to represent this variability more effectively and can help tighten the theoretical bound described in Theorem 1. However, using an excessive number of latent domains may weaken the intended benefit of merging difficult-to-align physical domains with intermediate ones, which is designed to alleviate manifold distortion. Moreover, this increases model fitting difficulty by reducing the number of samples available within each domain. It is also important to emphasize that our method remains grounded in Theorem 1 and Proposition 1. Its effectiveness depends on the assumption that the source and target domains share meaningful common components, even in the absence of explicit transformation functions or global physical priors. When the discrepancy between domains becomes excessively large, not only our approach but any domain generalization framework relying solely on source-domain data would inevitably fail. 
\section{\name System}

In this section, we provide a detailed introduction to the proposed adversarial generalization system for gesture recognition based on WiFi latent domain mining.

\subsection{System Overview}

\begin{figure}[htp]
  \centering
  \includegraphics[width=0.56\linewidth]{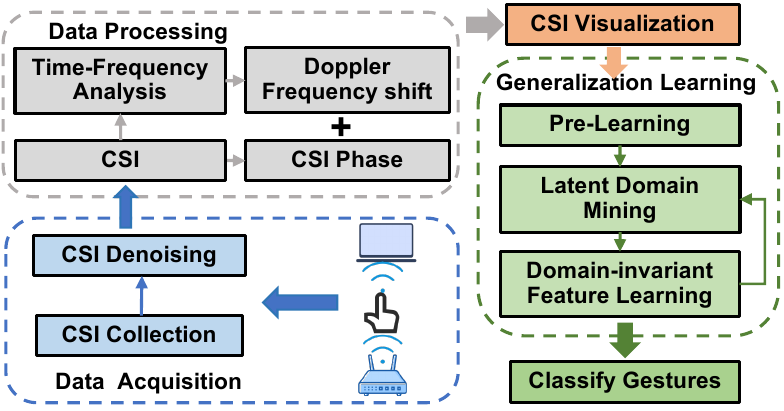}
  \caption{The system framework of \name.}
  \label{fig:sys}
\end{figure}

The system framework of \name~is illustrated in Figure \ref{fig:sys}. The framework comprises the following two components: data acquisition and processing, generalization learning. 

\textbf{Data Acquisition and Processing:} In the WiFi sensing scenario, the transmitter continuously sends WiFi signals, which are received by receivers. During this process, information about human gesture is recorded by CSI. CSI data are then fed into a denoising module, where the CSI ratio method~\cite{zeng2019farsense} is applied to suppress noise. Then, we apply the STFT to extract the DFS for frequency-domain features, while simultaneously retaining phase information as temporal-domain features. The extracted DFS and phase data are subsequently visualized into a unified image representation like UniFi~\cite{liu2024unifi}.

\textbf{Generalization Learning:} CSI images serve as the input of the  latent domain mining-based gesture recognition network. Initially, a network is roughly trained over two rounds on the gesture classification task to acquire preliminary knowledge of the data distribution. Following this, latent domains are iteratively discovered through clustering, combined with adversarial learning to suppress gesture-class-related information from the latent features. Finally, robust domain-invariant representations are learned by adversarially aligning the latent domains. The trained feature extractor, together with a softmax classifier, is then used for cross-domain gesture recognition.

\subsection{Data Acqusition and Processing}

This module is designed to establish a stable and reliable foundation for signal representation. The received CSI not only carries gesture-related information but is also significantly affected by hardware-induced phase noise $e^{-j\theta_{\text{n}}}$, which distorts the observed measurements
\begin{equation}
\label{equ:off}
H(f,t)= e^{-j\theta_{n}}(H_{s}(f,t)+ H_{d}(f,t)) = e^{-j\theta_{n}}(H_{s}(f,t)+ A(f,t) e^{-j2\pi \frac{d(t) }{\lambda } })
\end{equation}
where $A(f,t)$, $e^{-j2\pi \frac{d(t)}{\lambda}}$, and $d(t)$ represent the complex attenuation, the phase shift, and the propagation path length of the dynamic components, respectively. It is obvious that $e^{-j\theta_{\text{n}}}$ hinders the direct utilization of CSI phase.

Thus, it is necessary to first eliminate $e^{-j\theta_{n}}$. Fortunately, for commodity WiFi devices, $e^{-j\theta_{\text{n}}}$ remains constant across different antennas on the same WiFi Network Interface Card (NIC) because they share the same RF oscillator. Thus, $e^{-j\theta_{n}}$ can be eliminated using the CSI-ratio~\cite{wu2020fingerdraw}:
\begin{equation}
	\begin{aligned}
		\label{equ:csiratio}
		H_q(f,t) &= \frac{H_{1}(f,t)}{H_{2}(f,t)} 
		=\frac{e^{-j\theta _{n}}(H_{s,1}+A_{1}e^{-j2\pi \frac{d_{1}(t)}{\lambda } })}{e^{-j\theta _{n}}(H_{s,2}+A_{1}e^{-j2\pi \frac{d_{2}(t)}{\lambda } })}\\
		&=\frac{A_{1}e^{-j2\pi \frac{d_{1}(t)}{\lambda }} +H_{s,1}}{A_{2}e^{-j2\pi \frac{d_{1}(t) + Delta }{\lambda } } +H_{s,2}}
	\end{aligned}
\end{equation}
where $H_1(f,t)$ and $H_2(f,t)$ denote the CSI measurements from two adjacent antennas, and $\Delta d$ represents the difference in path length between them. When $H_{1}(f,t)$ and $H_{2}(f,t)$ are close to each other, $\Delta d$ can be regarded as a constant. According to Mobius transformation~\cite{zeng2019farsense}, equation \ref{equ:csiratio} represents transformations such as the scaling and rotation of the phase shift $e^{-j2\pi \frac{d_{1}(t)}{\lambda }}$ of antenna $1$ in the complex plane, and these transformations do not affect the changing trend of the CSI. 

The data processing pipeline involves extracting the DFS and phase information from CSI signals, followed by visualization, which is critical for providing high-quality input to the generalization learning module. Specifically, we first extract the phase information as:
\begin{equation}
    \mathcal{P}(f,t) = Angle(H_q(f,t))
\end{equation}

For DFS, previous approaches mostly relied on conjugate multiplication as a preprocessing step, while \name~replaced it with the CSI ratio. The advantage is that the CSI ratio represents transformations such as the rotation and scaling of CSI in the complex plane, thus avoiding negative effects introduced by $H_{2}(f,t)$. To further mitigate the cumulative error caused by $\Delta d$, we apply an antenna selection coefficient $s_a$ to select $H_{1}(f,t)$ and $H_{2}(f,t)$, calculated as follows:
\begin{equation}
    s_a = \frac{1}{C}\sum_{c=1}^{C}\frac{\text{var}(|H_a(f_c,t))|}{\text{mean}(|H_a(f_c,t)|)}   
\end{equation}
where $\text{var}(\cdot)$ and $\text{mean}(\cdot)$ denote the variance and mean value of amplitude readings for the $a$th antenna of the $c$th subcarrier. We select the antennas with the highest and lowest $s_a$ as $H_{1}(f,t)$ and $H_{2}(f,t)$, respectively. The rationale behind this is that CSI with larger variances is generally more sensitive to motion, while CSI with higher amplitude typically contains a larger static path component, making $H_{1(f,t)}$ less affected by $\Delta d$. To further suppress static path components without affecting the DFS extraction, we apply a high-pass filter to the CSI. Finally, the DFS is obtained by applying the STFT to the processed CSI.

For effective training, it is crucial that the network receives normalized and information-rich inputs. To achieve this, we adopt the CSI visualization technique proposed in~\cite{gu2022wigrunt}, which transforms CSI data into heatmaps while normalizing and suppressing background static components. Specifically, we independently visualize the phase and DFS signals for each selected antenna pair, generating two images per pair. These images are subsequently concatenated across antenna pairs to form a single input image. The reason for visualizing them first and then concatenating the images instead of concatenating the CSI matrices first and then visualizing it is that the former introduces background information contrast between different antenna pairs, which leads to background interference. 


\subsection{WiFi Latent Domain Mining Based Gesture Recognition Network}

\begin{figure*}[htp]
  \centering
  \includegraphics[width=0.95\linewidth]{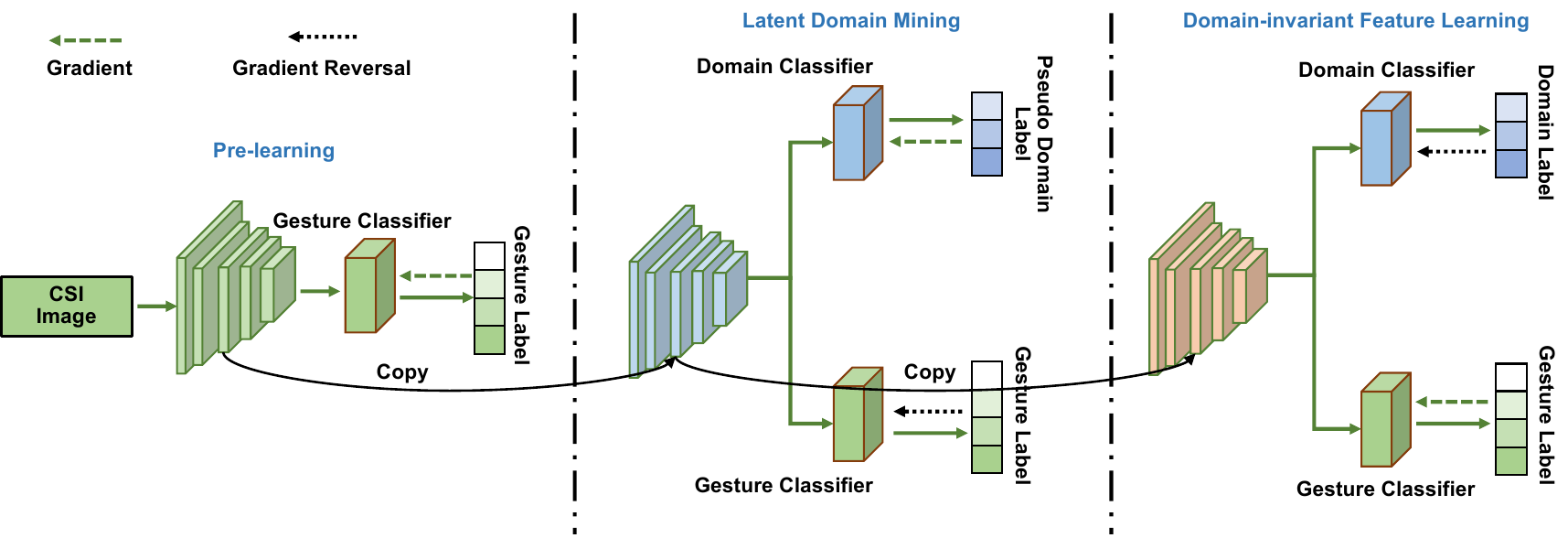}
  \caption{The WiFi latent domain mining based gesture recognition network.}
  \label{fig:net}
\end{figure*}

The proposed WiFi latent domain mining-based gesture recognition network, illustrated in Figure~\ref{fig:net}, comprises three steps, where Steps 2 and 3 are performed iteratively:
\begin{enumerate}
    \item Pre-learning: In this step, gesture labels are utilized as supervision to train the feature extraction module for several epochs, enabling the model to grasp relevant knowledge about the data distribution.
    \item Latent Domain Mining: The objective is to identify the latent domain labels of each sample based on the distribution knowledge already learned by the model and maximize the gaps between different domains.
    \item Domain-invariant Feature Learning: This step utilizes the latent-domain labels determined in the previous step to learn domain-invariant features and train a generalized gesture recognition model.
\end{enumerate}

\textbf{Pre-learning.} 
\name~aims to enable the model to autonomously uncover latent domains to avoid the distribution mismatch. Before characterizing the latent domains, we update the feature extractor module to equip it with knowledge relevant to data distribution. As shown in Figure \ref{fig:net} (green), where $h_f$, $h_b^p$, and $h_{c}^p$ represent the feature extractor, bottleneck, and gesture classifier, respectively. The cross-entropy loss is employed to compute the supervised loss:
\begin{equation}
    \mathcal{L}_{super} = \mathbb{L_C}(h_{c}^p(h_b^p(h_f(x))),y_{g})
\end{equation}
Where $x$ and $y_{g}$ represents input and label, respectively.

\textbf{Latent Domain Mining.} The purpose of this step is to uncover hidden distributions of the CSI data and to partition all samples into several distinctive latent domains. As illustrated in Figure \ref{fig:net} (gray), We employ a self-supervised pseudo-labeling strategy to assign latent domain labels. Specifically, we first assume that there are $k$ latent domains in the dataset and then compute the initial feature centroids for each domain:
\begin{equation}
    \tilde{\mu}_k = \frac{\textstyle \sum_{x\in \mathcal{X}^{tr} }^{}\delta_k(h_{c}^l(h_b^l(h_f(x))))h_b^l(h_f^l(x)) }{\textstyle \sum_{x\in \mathcal{X}^{tr}}^{}\delta_k(h_{c}^l(h_b^l(h_f(x))))}  
\end{equation}
Where $\mathcal{X}^{tr}$ is the distribution of training dataset, and $h_f$, $h_b^l$, and $h_c^l$ represent the feature extractor, bottleneck, and domain classifier, respectively. $\tilde{\mu}_k$ is the initial centroid of the $k$th latent domain, and $\delta_k$ is the $k$th logit softmax output of $h_{cd}^l$. 

Next, we assign pseudo-domain labels to all samples using the nearest-neighbor principle:
\begin{equation}
    y_d = arg \min_{k}\mathcal{D} (h_b^l(h_f(x)),\tilde{\mu}_k)
\end{equation}
Where $\mathcal{D}(a,b)$ is a distance measure function. Then, we iteratively compute the new domain feature centroids and pseudo-domain labels:
\begin{equation}
    \mu_k  = \frac{ {\textstyle \sum_{x \in \mathcal{X}^{tr} }^{} I(y_d  = k)h_b^l(h_f(x))} }{ {\textstyle \sum_{x \in \mathcal{X}^{tr}}^{}I(y_d = k)} }
\end{equation}
\begin{equation}
    y_d = arg \min_{k}\mathcal{D} (h_b^l(h_f(x)),\mu_k)
\end{equation}
Where $I(o) = 1$ if $o$ is true, otherwise 0. 

To ensure that the partitioning of latent domains is not influenced by gesture classification knowledge and to continuously enlarge the gap between distributions of different latent domains, the loss function for this step is as follows:
\begin{equation}
\label{eq:23}
	    \mathcal{L}_{lad} + \mathcal{L}_{adv} =
	     \mathbb{L_C}(h_c^l(h_b^l(h_f(x))),y_d) + \mathbb{L_C}(h_{adv}^l(\mathcal{R}_{\lambda_1}(h_b^l(h_f(x)))),y_g)
\end{equation}
Where $h_{adv}^l$ is the classifier that contains some linear layers and a softmax layer for classifying gestures. The $\mathcal{L}_{adv}$ suppresses gesture semantics, guiding the clustering process to better discover latent domains that capture the true distributional shifts. $\mathcal{R}_{\lambda_1}$ is the Gradient Reverse Layer (GRL) with hyperparameter $\lambda_1$. The GRL enables adversarial learning in a single forward–backward pass~\cite{ganin2015unsupervised}: during backpropagation, it multiplies the gradient from the domain discriminator by a negative constant, effectively reversing the gradient direction flowing into the feature extractor. The pseudo-domain labels $y_d$ obtained in this step are used for the subsequent domain-invariant gesture recognition feature learning.

\textbf{Domain-invariant Feature Learning.} After obtaining latent domain labels, we utilize adversarial learning to acquire generalizable features. As depicted in Figure \ref{fig:net} (orange), a gesture classifier and a domain adversarial classifier engage in a minimax game for feature learning, and the loss function is as follows:
\begin{equation}
    \mathcal{L}_{ges} + \mathcal{L}_{dadv} \\
    = \mathbb{L_C}(h_c^d(h_b^d(h_f(x))),y_g) + w_d\mathbb{L_C}(h_{dadv}(\mathcal{R}_{\lambda_2}h_b^d(h_f(x))),y_d)
\end{equation}
Among them, $w_d$ is the balancing weight, and it aims to prevent the imbalance of latent domain allocation from introducing bias in the final domain-agnostic feature learning step. By performing domain adversarial alignment using the mined latent domain labels $y_d$, rather than relying on physical domain labels, the model is better able to mitigate semantic conflicts and manifold distortion—thus facilitating the learning of more robust domain-invariant representations. The calculation method for $w_d$ is:
\begin{equation}
    w_d = \frac{ {\textstyle \sum_{d=1}^{K}}N_d }{N_d}
\end{equation}
Where $N_d$ is the number of samples in latent domain $d$. Regarding Eq~\ref{eq:23}, since we use gesture categories as the discriminator input and the dataset contains an equal number of samples per gesture class, no additional balancing coefficient is required.

The final loss is as follows:
\begin{equation}
    \mathcal{L} = \mathcal{L}_{super} + \mathcal{L}_{lad} + \mathcal{L}_{adv} + \mathcal{L}_{ges} + \mathcal{L}_{dadv}
\end{equation}
Although all loss terms are written together for compactness, it should be noted that the three steps are optimized step by step rather than jointly.
The last two steps are repeated until convergence or reach the maximum epoch. For gesture recognition, the final feature extractor $h_f(x)$ can be directly combined with the gesture classifier as follows:
\begin{equation}
    \hat{y}_g =h_c^d(h_b^d(h_f(x)))
\end{equation}

Since the feature extractor $h_f(\cdot)$ remains unchanged throughout the training process, the size of the feature extraction component in the final inference model is identical to that of the original backbone network. Moreover, both $h_b^d(\cdot)$ and $h_c^d(\cdot)$ are lightweight modules with a small number of parameters. As a result, the overall computational overhead during the inference phase increases only slightly compared to the original backbone network. In addition, the final model obtained through this process can be flexibly combined with other modules such as meta-learning, data augmentation techniques, or attention mechanisms to further enhance performance.
\section{Experiments}

\subsection{Datasets}

\begin{table*}[htp]
\footnotesize
\centering   	
\caption{Description of Widar3 dataset.}
\begin{tabular}{cm{0.8cm}<{\centering}m{7.5cm}<{\centering}m{1.3cm}<{\centering}m{1.3cm}<{\centering}m{1.2cm}<{\centering}}
    \toprule
    Environments & No. of Users & Gestures & No. of Locations &  No. of Orientations & No. of Samples\\
    \hline
    1 & 9 & 1: Push Pull; 2: Sweep; 3: Clap; 4:Slide; 5: Draw-O; 6: Draw-Zigzag; 7: Draw-N; 8: Draw-Triangle; 9: Draw-Rectangle; & 5 & 5 & 10125\\
    \hline
    2 & 4 & 1: Push Pull; 2: Sweep; 3: Clap; 4:Slide; 5: Draw-O; 6: Draw-Zigzag; & 5 & 5 & 3000\\
    \hline
    3 &  4 & 1: Push Pull; 2: Sweep; 3: Clap; 4:Slide; 5: Draw-O; 6: Draw-Zigzag; & 5 & 5 & 3000\\
    \bottomrule
\end{tabular}
\label{Widar3} 
\end{table*}

\begin{figure}[htp]
	\centering
		\subfloat[Widar 3.0]{\label{widar}
		\begin{minipage}{0.25\linewidth}
			\centering
			\includegraphics[width=1\textwidth]{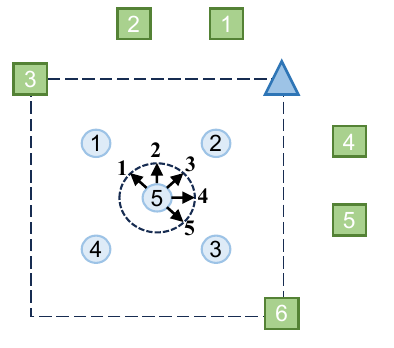}
		\end{minipage}
	}	
	\quad
	\subfloat[ARIL]{\label{aril}
		\begin{minipage}{0.29\linewidth}
			\centering
			\includegraphics[width=1\textwidth]{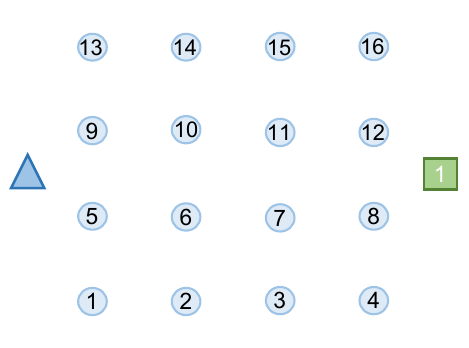}
		\end{minipage}
	}	
    \quad
	\subfloat[XRF55]{\label{xrf}
		\begin{minipage}{0.25\linewidth}
			\centering
			\includegraphics[width=1\textwidth]{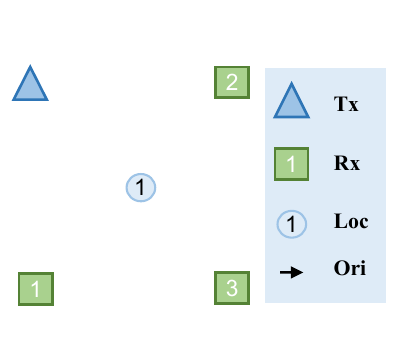}
		\end{minipage}
	}	
	\caption{Deployment Setting of Widar 3.0, ARIL and XRF55.}
	\label{fig:setting}
\end{figure}

In this section, we introduce the datasets used for experimental evaluation, including three public benchmarks, Widar3, ARIL and XRF55. Furthermore, to validate the effectiveness of \name, we conducted tests on a real-world gesture recognition deployment. Detailed descriptions of the real environment are provided in the corresponding section.

\textbf{Widar3} \cite{zhang2021widar3} contains $16,125$ samples collected from 3 environments, and its detailed description is shown in Table \ref{Widar3}. To evaluate the performance of~\name, we conduct in-domain and cross-location/user/orientation experiments using data from the first environment (6750 samples, 9 users $\times$ 5 positions $\times$ 5 orientations $\times$ 6 gestures $\times$ 5 instances). For cross-environment evaluation, we utilize data collected from all three environments. The deployment setup of Widar3 is illustrated in Figure~\ref{widar}.

\textbf{ARIL} \cite{wang2019joint} comprises $6$ distinct gestures (namely, hand up, hand down, hand left, hand right, hand circle, and hand cross), executed by a single user at $16$ distinct locations within a confined space. ARIL leverages universal software radio peripheral (USRP) devices for CSI data collection. The dataset comprises a total of $1,392$ samples.
The deployment setup of ARIL is shown in Figure\ref{aril}.

\textbf{XRF55}~\cite{wang2024xrf55} is a comprehensive multimodal dataset covering 55 types of indoor activities, incorporating data from WiFi devices, RFID tags, millimeter-wave radars, and visual sensing modalities. In our experiments, we select eight WiFi-measured gesture activities: drawing a circle, drawing a cross, pushing, pulling, swiping left, swiping right, swiping up, and swiping down. The XRF55 collection setup includes one transmitter and three receivers, each equipped with three antennas, with the transmitter configured to broadcast at 200 packets per second. The dataset involves 39 subjects performing the selected gestures across four distinct scenes. Each gesture is repeated 5 times within a 5-second window, resulting in a total of 6,240 gesture data samples. 

\subsection{Implementation Details} 

In our implementation, we conduct data acquisition and processing using Matlab, while the latent domain mining-based generalization gesture recognition network is constructed using the PyTorch framework. The number of latent domains is set to $3$. During training, we use an initial learning rate of $0.002$, which is reduced by a factor of $10$ every $10$ epochs for a total of $50$ epochs, with a batch size of $32$. In the pre-learning step, we train the feature extractor for $2$ epochs. We employ the Adam optimizer for our model, and the feature extraction backbone network is the ImageNet pre-trained ResNet18.
\subsection{Overall Performance}

\begin{table*}[ht]
\small
\centering   	
\caption{Results on Widar3 dataset with 6 antenna pairs. $^{\dagger}$ means domain adaption based method.}
\begin{tabular}{cccccccc}
    \toprule
    Methods & In-Domain & Cross-Loc & Cross-Ori  &  Cross-Env & Cross-User & Cross-Loc-Env & Cross-Ori-Env\\
    \hline
    EI~\cite{jiang2018towards} &97.40\% & 73.33\% & 79.70\% & 63.50\% & - & 20.73\% & 37.66\%\\
    Widar3.0~\cite{zhang2021widar3} & 79.25\% & 76.22\% & 78.07\% & 62.24\% & - & 53.17\% & 44.00\%\\
    WiHF~\cite{li2020wihf} &97.25\% & 89.11\% & 87.55\% & 82.25\% & -& 63.54\% & 77.33\%\\
    WiGRUNT~\cite{gu2022wigrunt} & 98.66\%& 92.07\% & 91.92\% & 85.19\% & - & 60.53\% & 73.00\%\\
    PAC-CSI~\cite{su2023real} &99.46\% & 98.77\% & 98.90\% & 96.47\% & 97.54\% & - & -\\
    UniFi~\cite{liu2024unifi} & 99.40\% & \underline{99.18\%} & \underline{99.40\%} & 97.73\% & 96.27\% & 95.65\% & \underline{95.33\%}\\
    MetaFormer$^{\dagger}$~\cite{sheng2024metaformer} & \textbf{100\%} & 99.00\% & \textbf{100\%} & 81.00\% & 94.67\% & - & -\\
    WiDUal~\cite{cao2025widual}& 99.67\% & 97.24\% & 94.91\% & - & - & - & -\\
    Wi-SFDAGR~\cite{yan2025wisfdar}& - & 97.30\% & 97.17\% & 95.52\% & - & - & - \\
    \name-Kmeans & 99.10\% & 99.11\% & 96.05\% & \underline{98.87\%} & \underline{98.98\%} & \underline{98.73\%} & \textbf{95.66\%}\\
    \name & \underline{99.73\%} & \textbf{99.26\%} & 96.45\% & \textbf{99.32\%} & \textbf{99.37\%} & \textbf{98.82\%} & \textbf{95.66\%}\\
    \bottomrule
\end{tabular}
\label{tab:overperw36d} 
\end{table*}

\begin{table*}[ht]
\small
\centering   	
\caption{Results on ARIL and Widar3 dataset with 1 antenna pair. $^{\dagger}$ means domain adaption based method and $^*$ denotes utilizes only two antenna pairs. (Note: Since the authors of UniFi provided results of $^*$ only in graphical form, the results we have filled in are approximations.)}
\begin{tabular}{ccccccc}
    \toprule
    Methods & In-Domain & Cross-Loc & Cross-Ori  &  Cross-Env &  Cross-User & ARIL-Cross-Loc\\
    \hline
    SelfReg~\cite{kim2021selfreg} & - & 76.71\% & 86.67\% & 39.11\% & 53.10\% & 44.45\%\\
    WiSGP~\cite{liu2023generalizing} & - &78.49\% & 88.46\% & 43.17\% & 56.77\% & 48.74\%\\
    WiSR~\cite{liu2023wisr} & - &77.51\%& \underline{88.80\%} & 42.52\% & 55.18\% & 48.64\%\\
    UniFi$^*$~\cite{liu2024unifi} & \textbf{97.50\%} & \underline{92.50\%} & \textbf{92.00\%} & \underline{87.50\%} & - & -\\
    Wi-Learner$^{\dagger}$~\cite{feng2022wi} & 93.20\% & 91.40\% & 86.50\% & 74.20\% & \underline{89.40\%} & -\\
    WiOpen~\cite{zhang2025wiopen} & - & 86.40\% & 77.67\% & 84.44\% & 82.71\% & \underline{73.61\%}\\
    \name & \underline{97.29\%} & \textbf{95.01\%} & 86.66\% & \textbf{93.20\%} & \textbf{93.67\%} & \textbf{75.52\%}\\
    \bottomrule
\end{tabular}
\label{tab:overperwrid} 
\end{table*}

\begin{table}[tp]
\small
\centering   	
\caption{Results on XRF55 dataset.}
\begin{tabular}{cccc}
    \toprule
    Methods & In-Domain & Cross-Env & Cross-Usr\\
    \hline
    ImgFi~\cite{zhang2023imgfi} & 39.13\% & 31.90\% & 33.39\%\\
    THAT~\cite{li2021two} & 76.25\% & 23.23\% & 31.77\%\\
    WiSR~\cite{liu2023wisr} & 67.53\% & 26.66\% & 32.72\%\\
    WiGRUNT~\cite{gu2022wigrunt} & \underline{90.48\%} & 55.92\% & \underline{63.47\%}\\
    Wi-SFDAGR~\cite{yan2025wisfdar}& - & \underline{57.99\%} & -\\
    \name & \textbf{92.00\%} & \textbf{62.15\%} & \textbf{67.18\%} \\
    \bottomrule
\end{tabular}
\label{tab:xrf55} 
\end{table}

The overall performance of~\name~is summarized in Tables~\ref{tab:overperw36d}, \ref{tab:overperwrid}, and \ref{tab:xrf55}, which compare~\name~against SOTA solutions on both in-domain and cross-domain tasks. It is evident that~\name~consistently outperforms existing domain generalization gesture recognition approaches across the majority of tasks, and even surpasses domain adaptation-based methods (require access to target domain data) in several scenarios. Notably, compared to domain adversarial-based generalization approaches using physical domain label~\cite{jiang2018towards,liu2023wisr}, \name~achieves substantially greater performance improvements, effectively demonstrating the effectiveness of the proposed WiFi latent domain mining strategy, thereby unlocking the potential of adversarial learning.


For Widar3 with a single antenna pair, as well as the ARIL dataset (which inherently uses a single antenna pair), \name~shows relatively lower performance compared to certain domain generalization and domain adaptation baselines in the in-domain and cross-orientation tasks, with performance gaps ranging from 1\% to 6\%. Nevertheless, in other cross-domain settings, \name~maintains significant advantages. For instance, it outperforms WiSGP~\cite{liu2023generalizing} by over 36\% and 26\% in the cross-user task and the ARIL cross-location task, respectively, and even exceeds the domain adaptation solution Wi-Learner~\cite{feng2022wi} by more than 18\% in the cross-environment task. Furthermore, \name~achieves gains of over 2\% and 5\% against UniFi in cross-location and cross-environment evaluations, despite UniFi utilizing data from two antenna pairs.



For the XRF55 dataset, where domain shifts are primarily caused by variations in users and environments, we focused our evaluation on cross-user and cross-environment scenarios.
Since XRF55 is a newly released dataset with no prior published benchmarks (only Wi-SFDAGR~\cite{yan2025wisfdar}), we compared \name~against several open-source baseline methods, as shown in Table~\ref{tab:xrf55}. The Widar3 method was excluded due to its high dependency on CSI data quality, making it unsuitable for XRF55 conditions. It can be observed that \name demonstrates substantial improvements over existing methods. The performance degradation observed on XRF55 is considerably greater than that on Widar3, with accuracy dropping by approximately 23–30\% compared to the within-domain results. We attribute this discrepancy primarily to the substantial differences in data quantity, which plays a crucial role in training neural networks for cross-domain scenarios. For instance, under the cross-environment setting, the Widar dataset contains 12,750 samples, averaging about 2,550 samples per gesture class, whereas XRF55 provides only around 780 samples per gesture. Furthermore, the data quality of the XRF55 and ARIL is relatively lower, introducing additional noise and variability that further exacerbate the generalization challenge.

\subsection{Cross-domain Study}

\begin{figure*}[htp]
	\centering
		\subfloat[Cross Location]{\label{cl}
		\begin{minipage}{0.45\linewidth}
			\centering
			\includegraphics[width=1\textwidth]{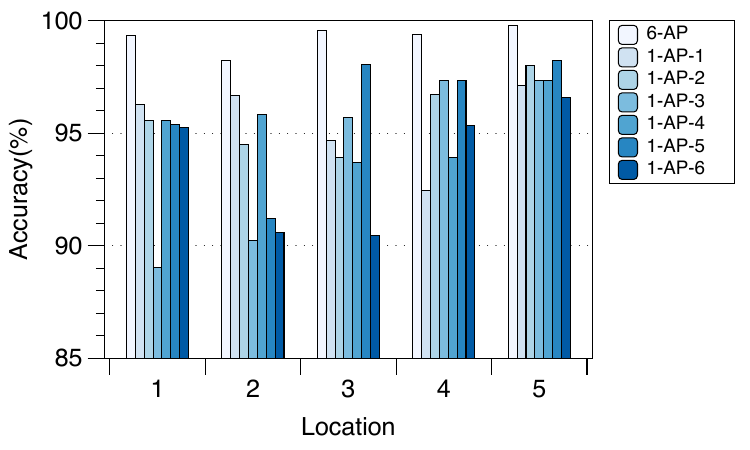}
		\end{minipage}
	}	
	\quad
	\subfloat[Cross Orientation]{\label{co}
		\begin{minipage}{0.45\linewidth}
			\centering
			\includegraphics[width=1\textwidth]{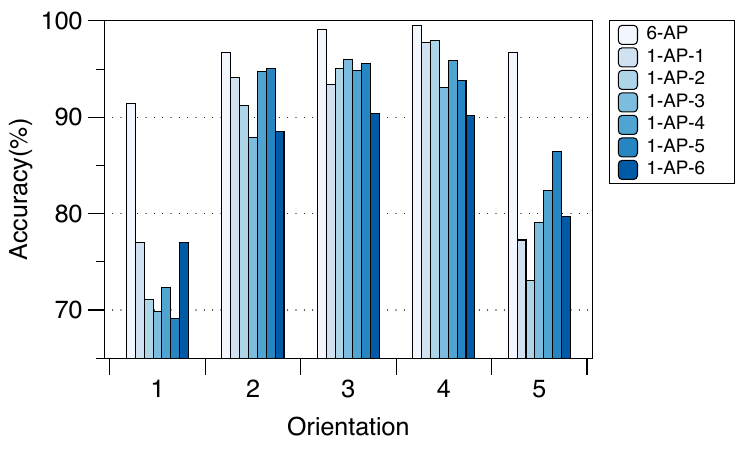}
		\end{minipage}
	}	
	
	\quad
        \subfloat[Cross Environment]{\label{ce}
		\begin{minipage}{0.45\linewidth}
			\centering
			\includegraphics[width=1\textwidth]{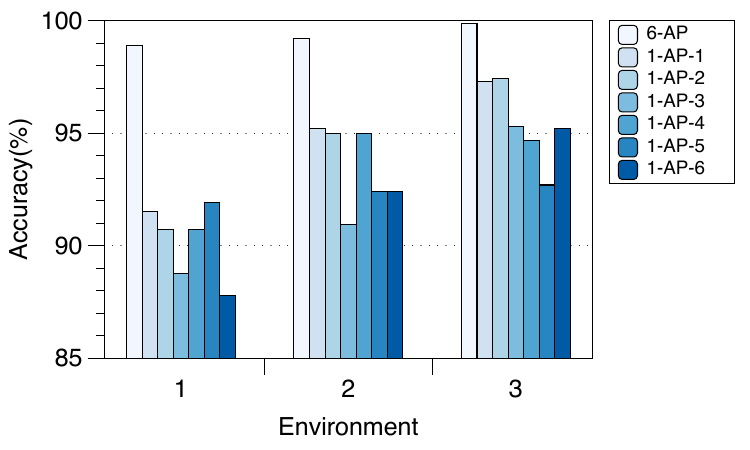}
		\end{minipage}
	}	
	\quad
        \subfloat[Cross Location ARIL]{\label{ceo}
		\begin{minipage}{0.45\linewidth}
			\centering
			\includegraphics[width=1\textwidth]{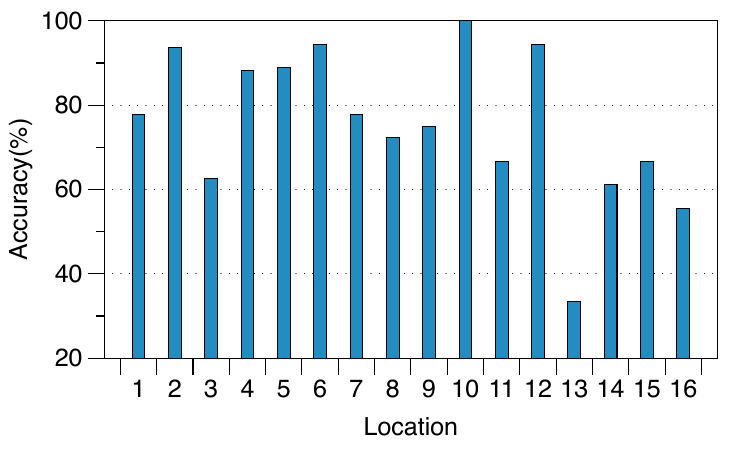}
		\end{minipage}
	}	
	\quad
	\caption{Cross Single Domain Performance.}
	\label{fig:cdres}
	\vspace{-0.16in}
\end{figure*}

\begin{figure}[tp]
	\centering
        \subfloat[Cross Location Environment]{\label{cel}
		\begin{minipage}{0.46\linewidth}
			\centering
			\includegraphics[width=1\textwidth]{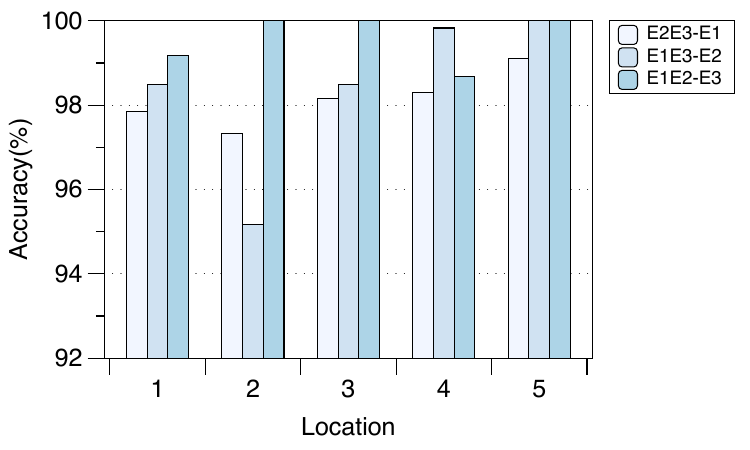}
		\end{minipage}
	}	
	\quad
        \subfloat[Cross Orientation Environment]{\label{cla}
		\begin{minipage}{0.46\linewidth}
			\centering
			\includegraphics[width=1\textwidth]{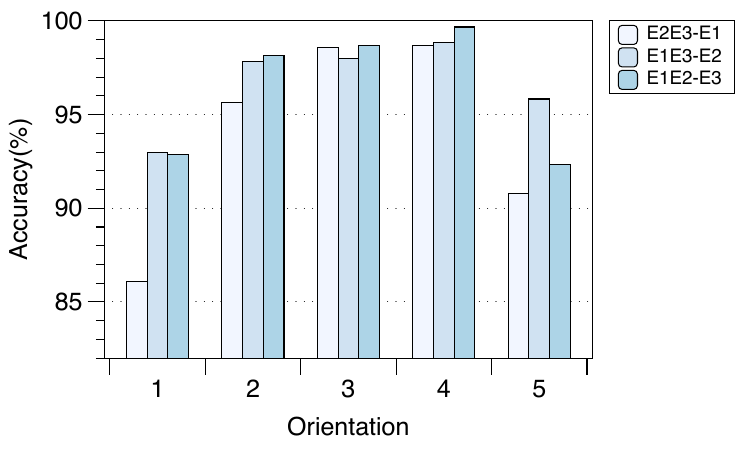}
		\end{minipage}
	}	
	\quad
	\caption{Cross Multiple Domain Performance.}
	\label{fig:c2dres}
\end{figure}


We present experimental results for selected cross single-domain tasks, as shown in Figure~\ref{fig:cdres}. It can be observed that gesture recognition performance exhibits strong correlations with domain factors, and these correlations cannot be inferred solely from physical domain labels. In Figure~\ref{cl}, for the cross-location task, except for antenna pairs 1 and 4 at location 2, the performance of other single antenna pairs is relatively poor. In Figure~\ref{co}, for the cross-orientation task, orientations 1 and 5 exhibit the worst performance, whereas orientations 2, 3, and 4 achieve better results. We attribute this to the relative positioning between antenna pairs and users, as well as the degree of domain continuity. As depicted in Figure~\ref{widar}, being either too close to (e.g., antenna pairs 3 and 6 at location 2) or too far from (e.g., antenna pair 1 at location 4) the antennas can reduce sensing sensitivity. In the cross-location task, location 5 achieves the best performance because it is surrounded by the other four locations, thus benefiting from more continuous domain variation data. In contrast, in the cross-orientation task, orientations 1 and 5 perform the worst, as they are located at the endpoints of the orientation domain variation, while orientations 2, 3, and 4 have more intermediate domain data to facilitate generalization. Furthermore, compared to the results in Figure~\ref{fig:grlanddg}, \name achieves a substantial performance improvement.


In the cross-environment task, poor performance in environment 1 can be attributed to the limited diversity of training data from environments 2 and 3. The subpar performance in environment 2 is due to its data being collected over five days, resulting in a larger temporal span compared to the other environments, where data collection was completed within a single day. In the ARIL dataset, performance also exhibits strong location dependency. However, the overall lower data quality increases uncertainty factors, further highlighting the inadequacy of using subjective physical domain labels. The performance across cross-multidomain tasks is shown in Figure~\ref{fig:c2dres}, revealing variations that correlate with both task types and antenna perspectives. Despite the complexity of cross-multidomain tasks, \name~continues to deliver robust performance, underscoring the effectiveness of WiFi latent domain exploration strategies.

To further examine whether the automatically discovered latent domains correspond to specific physical factors, we analyzed the composition of these factors within each mined latent domain using the results from the cross-orientation and cross-location tasks. For each latent domain $k$, we computed the purity and entropy of several physical attributes, including user, position, and orientation. The purity is calculated as follows:
\begin{equation}
\text{Purity}_k =
\begin{cases}
\dfrac{\max_i \text{cnt}_i}{\sum_j \text{cnt}_j}, & \text{if } \sum_j \text{cnt}_j > 0 \\
0, & \text{otherwise}
\end{cases}
\end{equation}
Here, $\text{cnt}_i$ denotes the number of samples in the $k$-th latent domain whose target physical factor takes the label $i$. This metric reflects whether the latent domain is dominated by a single factor label. A higher purity indicates that the domain is more concentrated or “pure” with respect to that factor, while a lower value suggests greater mixing. The entropy is defined as follows:
\begin{equation}
p_i = \dfrac{c_i}{\sum_j c_j}, \quad
H_k = -\sum_i p_i \ln \left(p_i + 1\times10^{-12}\right)
\end{equation}
$H_k=0$ means that the latent domain is almost entirely composed of a single factor value (extremely pure), whereas $H_k\approx ln^M$ (where $M$ is the number of possible labels for that factor) indicates that the distribution is close to uniform, implying that the latent domain is independent of this factor.

As shown in Table~\ref{tab:latent_purity_entropy}, all factors exhibit low purity and high entropy values close to those of a theoretical uniform distribution. This indicates that each latent domain does not correspond to a single physical factor. Instead, the latent domains appear to capture compound CSI distributions influenced by the combined effects of multiple physical variations, forming abstract statistical regimes that are well-suited for adversarial generalization rather than explicit physical partitions.

\begin{table}[t]
\small
\centering
\caption{Purity and entropy of different physical factors across three cross-directional tasks. Theoretical values for a perfectly uniform distribution are provided for reference.}
\begin{tabular}{lcccccccc}
\toprule
\multirow{2}{*}{Factor} & 
\multicolumn{2}{c}{Purity ($\uparrow$)} & & 
\multicolumn{2}{c}{Entropy ($\downarrow$)} & &
\multirow{2}{*}{Ideal Purity} & 
\multirow{2}{*}{Ideal Entropy} \\
\cmidrule{2-3} \cmidrule{5-6}
& Cross-ori & Cross-loc & & Cross-ori & Cross-loc & & & \\
\midrule
 User& 0.142 & 0.147  & & 2.182 & 2.173 & & 0.111 & 2.197 \\
 Location & 0.226 & 0.270 & & 1.606 & 1.384 & & 0.200/0.250 & 1.609/1.386 \\
 Orientation & 0.295 & 0.216  & & 1.375 & 1.606 & & 0.250/0.200 & 1.386/1.609 \\
 Loc+Ori & 0.073 & 0.063  & & 2.966 & 2.981 & & 0.050 & 2.996 \\
 User+Loc+Ori & 0.012 & 0.011 & & 5.042 & 5.105 & & 0.006 & 5.193 \\
\bottomrule
\end{tabular}
\label{tab:latent_purity_entropy}
\end{table}

\subsection{Ablation Study}


\begin{figure*}[htp]
	\centering
        \subfloat[Number of Latent Domains]{\label{nolado}
		\begin{minipage}{0.39\linewidth}
			\centering
			\includegraphics[width=1\textwidth]{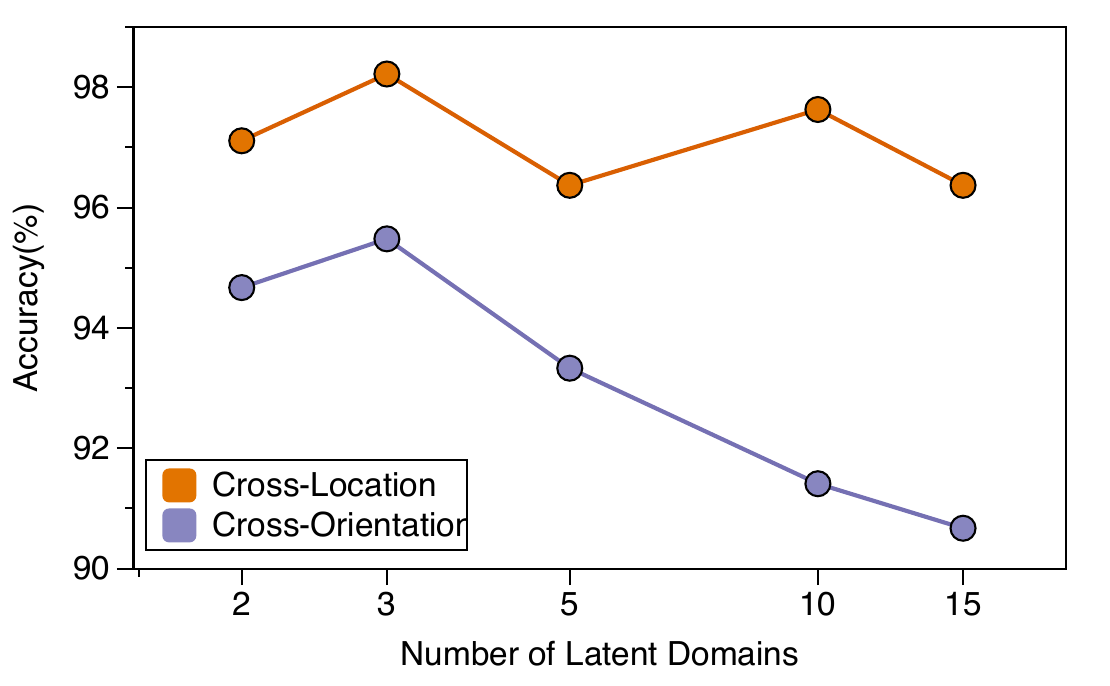}
		\end{minipage}
	}	
	\quad
        \subfloat[Adversarial Gesture]{\label{gesadv}
		\begin{minipage}{0.25\linewidth}
			\centering
			\includegraphics[width=1\textwidth]{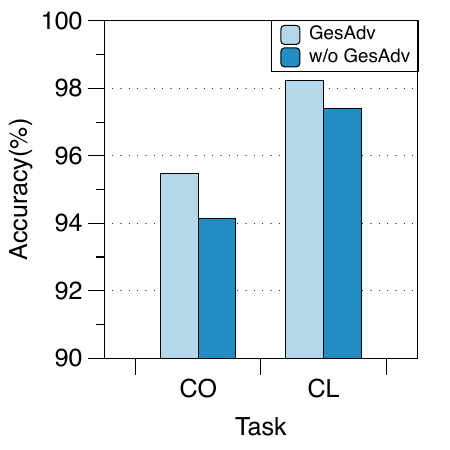}
		\end{minipage}
	}	
	\quad
        \subfloat[Domain Weighting]{\label{dowei}
		\begin{minipage}{0.25\linewidth}
			\centering
			\includegraphics[width=1\textwidth]{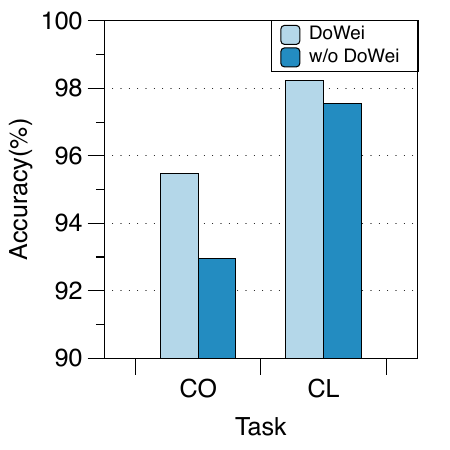}
		\end{minipage}
	}	
	\quad
	\caption{Ablation Study.}
	\label{fig:ablat}
    \vspace{-0.0in}
\end{figure*}

\begin{table}[t]
\small
\centering
\caption{Accuracy under different input modalities.
DP denotes \textit{DFS + Phase}, and DPA denotes \textit{DFS + Amplitude + Phase}.}
\label{tab:input_modalities_accuracy}
\begin{tabular}{lccccc}
\toprule
Setting & Amplitude & Phase & DFS & DPA & DP \\
\midrule
Cross-Orientation & 92.81\% & 92.96\% & 93.85\% & 95.04\% & \textbf{96.45\%} \\
Cross-Location  & 96.88\% & 97.04\% & 98.15\% & 97.70\% & \textbf{99.26\%} \\
\bottomrule
\end{tabular}
\end{table}

We conduct ablation experiments to evaluate the effectiveness of each module, using cross-location (with location 2 as the test set) and cross-orientation (with orientation 5 as the test set) tasks on the Widar3 dataset. The results are presented in Figure~\ref{fig:ablat}. It is evident that the best performance is achieved when the number of latent domains is set to 3. The default setting of three latent domains represents a practical balance between capturing the diversity of the data distribution and maintaining efficient training. Using too few latent domains may fail to adequately model the variability inherent in the data, while using too many may diminish the intended effect of merging difficult-to-align physical domains with intermediate ones, which is designed to alleviate manifold distortion.  Furthermore, increasing the number of latent domains reduces the number of samples available within each domain and exacerbates data imbalance, thereby increasing the difficulty of model optimization. In particular, when the number of latent domains exceeds that of the physical domains, it may even exacerbate distortion in challenging tasks such as cross-orientation. The gesture adversarial module introduced in Step 2 also demonstrates a positive impact, as it helps avoid the introduction of gesture related biases during latent domain partitioning. Furthermore, the domain weighting scheme employed in Step 3 effectively prevents domains with larger sample sizes from dominating feature learning, thereby preserving recognition performance in domains with fewer samples.

We further examined the influence of high-quality input representations on the performance of \name. Table~\ref{tab:input_modalities_accuracy} presents the results under different input modalities. Consistent with the observations reported in UniFi~\cite{liu2024unifi}, both DFS and phase individually provide strong performance, while their combination (DP) yields the best generalization capability across domains. Incorporating amplitude (DPA) leads to a slight reduction in accuracy due to its sensitivity to static background reflections, which is also consistent with UniFi’s findings. These results validate that the physically grounded DFS and Phase feature combination forms a robust and domain-generalizable representation for WiFi sensing~\cite{liu2024unifi}, effectively complementing our proposed latent-domain learning framework. To further evaluate the effectiveness of the proposed clustering strategy, we replaced the latent domain mining module in \name with a standard K-means algorithm while keeping the rest of the framework unchanged. The resulting accuracies are summarized in Table~\ref{tab:overperw36d}. Overall, K-means achieves competitive performance, with only marginally lower accuracy across all settings, confirming the robustness of our latent domain mining design. The key distinction between our approach and K-means lies in the centroid assignment process. Our method adopts a softmax-based soft centroid allocation, which aligns centroids with sensing-consistent statistical characteristics, provides a more informative initialization for clustering, and yields slightly higher as well as more stable performance.

\subsection{Impact of the Amount and Diversity of Training Data}
\begin{figure}[tp]
	\centering
        \subfloat[Cross Location]{\label{numcl}
		\begin{minipage}{0.45\linewidth}
			\centering
			\includegraphics[width=1\textwidth]{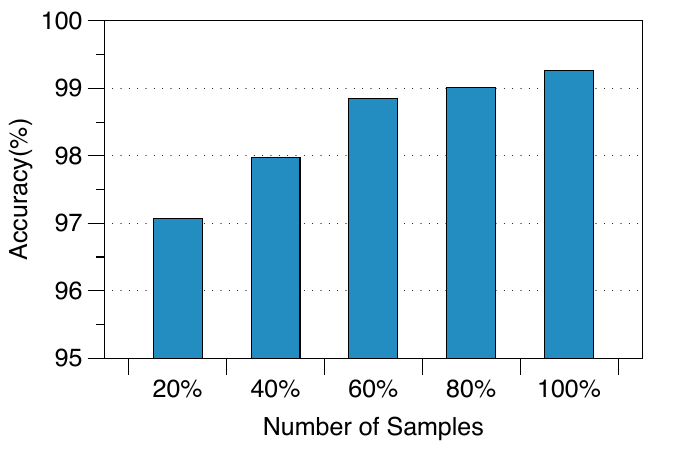}
		\end{minipage}

 }	
	\quad
        \subfloat[Cross Orientation]{\label{numco}
		\begin{minipage}{0.45\linewidth}
			\centering
			\includegraphics[width=1\textwidth]{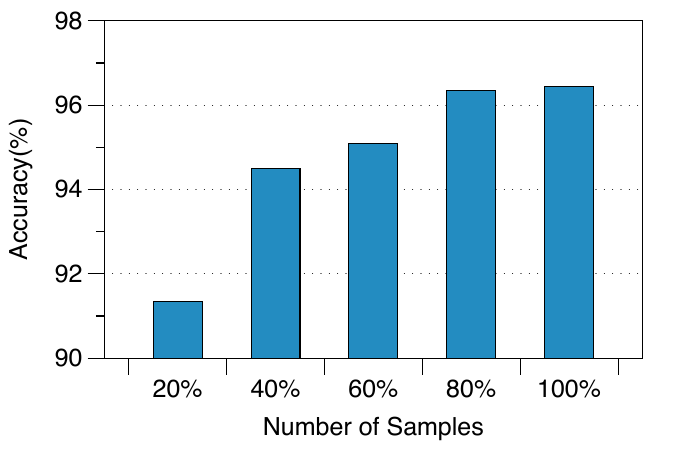}
		\end{minipage}
	}	
	\quad
	\caption{Influence of different amounts of training data.}
	\label{fig:numsam}
\end{figure}

We examine how the number of samples affects \name's performance by conducting experiments on cross-orientation and cross-location tasks using the Widar3 dataset. The results are illustrated in Figure \ref{fig:numsam}. It is evident that reducing the size of the training set to 80\% of its original size has a negligible impact on performance. \name~maintains good performance even when only 20\% of the samples are used for training. Additionally, simpler tasks, such as cross-location, are less affected by the reduction in sample size.

\begin{table}[t]
\centering
\small
\caption{Average accuracy under different numbers of training domains for cross-location and cross-orientation settings.}
\label{tab:training_diversity_avg}
\begin{tabular}{lccc}
\toprule
Setting & 1 Domain & 2 Domains & 3 Domains \\
\midrule
Cross-Location   & 95.36\% & 98.30\% & 99.85\% \\
Cross-Orientation & 66.35\% & 82.46\% & 92.60\% \\
\bottomrule
\end{tabular}
\end{table}

To disentangle the influence of data diversity, we varied the number of source domains (one, two, and three) and evaluated the model on an unseen target domain. As shown in Table~\ref{tab:training_diversity_avg}, \name maintains strong performance in the cross-location setting even with a single source domain, and its accuracy nearly saturates as domain diversity increases. In contrast, the cross-orientation scenario is far more sensitive to diversity: accuracy drops to 66.35\% when trained on a single source orientation but recovers to 92.60\% when trained on three. This observation can be attributed to the fact that orientation changes induce stronger distributional shifts in CSI data. Prior research has reported that gestures performed in opposite directions can produce nearly inverted CSI waveforms~\cite{niu2021understanding,niu2018fresnel}. Without explicit physical priors to describe how orientation affects CSI waveform transformations, a purely data-driven model trained on limited orientation domains struggles to infer or compensate for such shifts. Consequently, the latent domain partitioning also fails to capture transferable and robust representations. This limitation is not unique to \name but is inherent to data-driven approaches in general. Therefore, for WiFi sensing tasks where collecting sufficiently diverse datasets is challenging, we believe that integrating data-driven learning with physics-informed modeling offers a promising direction for developing the next generation of generalizable WiFi sensing systems.

\subsection{Experiments in Real-World Environments and Cross-Dataset Evaluation}

\begin{table}[htp]
\small
\centering 
\caption{Results on real-world environment and cross-dataset evaluation.}
\begin{tabular}{cccccc}
    \toprule
    Methods & In-Domain & Cross-Loc & Cross-Ori & Cross-User & Cross-Dataset\\
    \hline
    ImgFi \cite{zhang2023imgfi} & 24.67\% & 21.11\% & 22.00\% & 21.11\% & 25.46\%\\
    THAT \cite{li2021two} & 41.33\% & 28.89\% & 31.33\% & 29.11\% & 25.19\%\\
    WiSR \cite{liu2023wisr} & 42.67\% & 25.56\% & 27.33\% & 24.89\% & 26.67\%\\
    WiGRUNT \cite{gu2022wigrunt} & \underline{54.44\%} & \underline{38.89\%} & \underline{38.00\%} & \underline{38.22\%} & \underline{32.67\%}\\
    \name & \textbf{62.89\%} & \textbf{46.00\%} & \textbf{42.89\%} & \textbf{45.56\%} & \textbf{39.67\%}\\
    \bottomrule
\end{tabular}
\label{tab:rwe}
\end{table}

We also evaluated the performance of \name~in a more challenging real-world scenario. In a larger space, we deployed one transmitter and three receivers, where the transmitter was equipped with one antenna and each receiver with three antennas. Two users performed five gestures at five different positions and in three different orientations, with each gesture repeated three times, resulting in a total of 450 samples. The gestures included Push \& Pull, Clap, Slide, Slide Left, Slide Right, and Draw O. The experimental results are presented in Table~\ref{tab:rwe}. It is evident that \name~significantly outperforms all baseline methods, demonstrating the effectiveness of the proposed framework. During data collection, the distance between transceivers was notably larger than that used in datasets such as Widar3, and no restrictions were imposed on surrounding human activities. Consequently, frequent movements of nearby individuals introduced substantial dynamic multipath interference, which degraded the quality of the CSI signals. These factors collectively made the experimental environment considerably more complex and realistic, resulting in overall accuracy that was lower than that achieved on other datasets.

We further conducted a cross-dataset experiment in which \name was trained on the Widar3 dataset and evaluated on our real-world collected dataset. In our dataset, gestures were performed vertically, whereas in Widar3 they were horizontal. The comparative results with several representative baselines are reported in Table~\ref{tab:rwe}. Although the overall accuracy is lower than that of intra-dataset evaluations due to inevitable differences in hardware configurations, background noise, and gesture semantics, \name consistently outperforms all competing methods. This demonstrates its superior capability to learn transferable and domain-invariant representations.

\section{Discussion}

In this section, we discuss some limitations and potential improvements.

\begin{figure}[tp]
	\centering
        \subfloat[Cross Location]{\label{cdcol}
		\begin{minipage}{0.45\linewidth}
			\centering
			\includegraphics[width=1\textwidth]{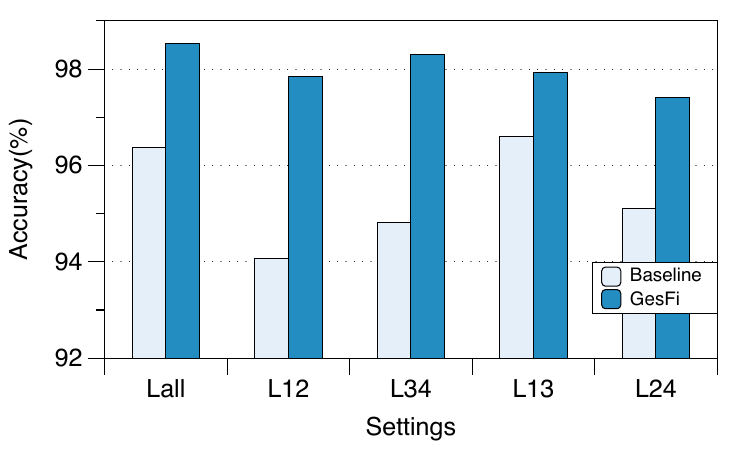}
		\end{minipage}
	}	
	\quad
        \subfloat[Cross Orientation]{\label{cdcoo}
		\begin{minipage}{0.45\linewidth}
			\centering
			\includegraphics[width=1\textwidth]{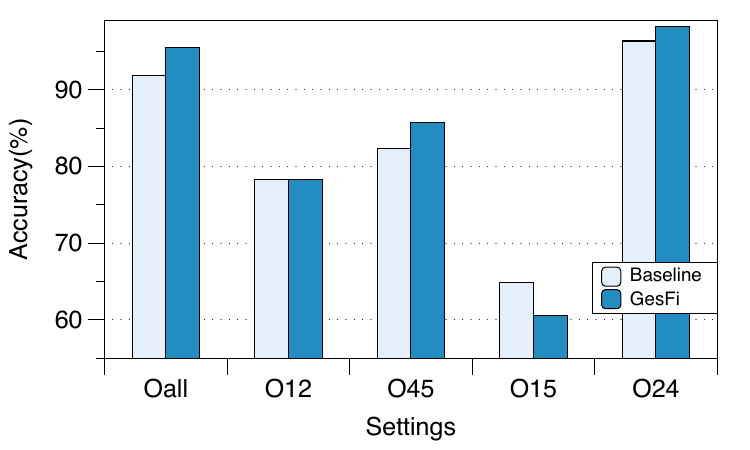}
		\end{minipage}
	}	
	\quad
	\caption{Influence of Domain Continuity Variations.}
	\label{fig:CdCo}
\end{figure}

\textbf{Improve latent domain mining in WiFi sensing data.} Currently, \name~considers the number of latent domains as a tunable hyperparameter. However, the optimal number of latent domains may vary depending on the distribution of WiFi sensing data encountered. Therefore, developing a method to automatically determine the optimal number of latent domains would be highly beneficial. We propose that the distinctions among latent domains in WiFi-based recognition are related to the semantics of gestures and the perspectives of antenna pairs observing users. Thus, exploring how to integrate contextual and expert knowledge to assist the model in automatically determining the appropriate number of latent domains is a promising direction for future research. Additionally, such an investigation could provide insights into the relationship between domain disparities in WiFi-based recognition and these factors, thereby advancing research toward more practical applications. For automatically determine the optimal number of latent domains, one strategy is to begin with a larger initial number and apply a series of regularization and refinement steps to adaptively reduce it. This includes: Regularizing the assignment entropy to encourage compact domain assignments, pruning low-mass domains with very few samples, and merging domains with highly similar feature distributions. Another promising direction is to adopt a meta-learning approach, in which a separate controller network learns to infer the optimal number of latent domains from the data distribution itself. We plan to incorporate such mechanisms in future work.

\textbf{Exploring true physical domain transition manifolds.} To investigate the limitation of \name, we conducted additional experiments on the Widar3.0 dataset, focusing on cross-location and cross-orientation tasks centered at location 5 and orientation 3. To ensure fairness across experimental settings, we controlled for sample size by utilizing only 20\% of the dataset when training with data from four positions. The results are shown in Figure~\ref{fig:CdCo}. An intriguing observation is that training solely with data from orientations 2 and 4 yields better generalization performance than using data from all other orientations, highlighting the existence of manifold distortion. Moreover, it is important to note that \name~trained with data from orientations 1 and 5 shows inferior performance compared to the baseline. This degradation occurs because, without intermediate transition physical domains, \name~also struggles to overcome the manifold distortion induced by forced domain alignment. These findings indicate that while~\name can partially address the biases introduced by physical domains (cases such as L12, L34, and performance compared to Figure~\ref{fig:grlanddg}), it still faces significant challenges when discrepancies between source domains become too substantial. Moving forward, incorporating physical models and leveraging the positional information of transceivers and users during sensing could offer promising avenues to guide domain alignment more effectively and further mitigate manifold distortion. Examples include differentiable channel models, neural field style scene representations that encode transceiver geometry and materials, and pose conditioned forward models that predict how a gesture maps to CSI under different orientations and locations.

\textbf{Remaining Challenges in Generalized WiFi Recognition.} Although our approach has made some progress toward achieving generalized WiFi-based gesture recognition, several key challenges remain. First, when antenna placements or gesture execution settings differ substantially, the bounded-divergence assumption no longer holds, leading to inevitable accuracy degradation. A purely data-driven model cannot readily infer the transformation that maps CSI distributions from a source domain to a markedly different target domain. We believe that integrating data-driven learning with physics-informed modeling represents a promising direction for developing the next generation of generalizable WiFi sensing systems. Second, unlike in the vision domain, collecting sufficiently diverse and extensive WiFi datasets remains extremely challenging. Nevertheless, the relatively low dimensionality of WiFi signals makes theoretically grounded simulation and synthetic data generation more tractable. A promising method is to leverage large language models to generate realistic environmental descriptions and subsequently, based on the generated environmental context, noise characteristics, and gesture semantics, synthesize CSI data grounded in wireless propagation theory. Such synthetic data could substantially enhance domain diversity and improve recognition system robustness. Third, variations in NIC drivers, phase sanitization, and clock offsets often introduce instability in DFS and phase statistics, thereby limiting the portability of trained models across devices. Addressing this issue requires stronger normalization pipelines, self-supervised calibration mechanisms, and lightweight test-time statistical alignment strategies. Finally, in real-world deployments, the diversity of gestures and activities encountered far exceeds those represented in existing datasets. Effectively filtering out irrelevant motions while reliably recognizing target gestures remains an open yet practically significant challenge.

\section{Related Work}

The evolution of WiFi-based gesture recognition has progressed from leveraging Received Signal Strength (RSS) to exploiting fine-grained CSI, and from handcrafted feature-based designs to deep learning-driven frameworks.

\textbf{Handcrafted Feature-based Approaches.}
Early studies primarily relied on manually engineered features extracted from RSS or CSI measurements. WiGest~\cite{abdelnasser2015wigest} modeled each gesture using handcrafted RSS patterns and performed recognition through similarity matching. Although innovative, its reliance on coarse-grained RSS metrics constrained accuracy. WiMU~\cite{venkatnarayan2018multi} improved recognition by exploiting fine-grained CSI for multi-user scenarios, yet its exhaustive search process for predefined gestures limited scalability. WiDraw~\cite{sun2015widraw} utilized Angle of Arrival (AoA) measurements for trajectory tracking but required more than twenty-five WiFi transceivers, making it impractical for real-world deployment. Similarly, QGesture~\cite{yu2018qgesture} achieved comparable accuracy using CSI phase information but required prior knowledge of the initial hand position. These approaches demonstrated the feasibility of WiFi-based sensing; however, handcrafted features must be specifically designed for each individual task and inevitably fail to capture all task-relevant characteristics.

\textbf{Machine Learning-based Methods.}
To overcome the limitations of handcrafted-based approaches, learning-based approaches sought to automatically extract discriminative representations directly from data. Wikey~\cite{ali2015keystroke} and WiFinger~\cite{li2016wifinger} were early examples that combined handcrafted CSI features with shallow classifiers such as SVM and KNN for keystroke recognition. Then, the introduction of deep learning frameworks significantly enhanced recognition capability. WiSign~\cite{zhang2020wisign} employed Deep Belief Networks (DBN) to process CSI amplitude and phase information for American Sign Language recognition. Widar 3.0~\cite{zhang2021widar3} proposed the Body-coordinate Velocity Profile (BVP), a domain-agnostic feature that captures the distribution of signal power over velocities, and integrated it with neural networks to improve cross-domain robustness. WiHF~\cite{li2020wihf} further developed domain-agnostic motion pattern features and integrated them with deep neural networks to achieve generalized gesture recognition. Nevertheless, these methods still partly relied on manually designed feature templates and overlooked the full representational capacity of end-to-end neural networks.

\textbf{Cross-domain WiFi-based Gesture Recognition.}
With the increasing demand for robust performance across users, locations, and environments, recent research has shifted toward domain adaptation and domain generalization. WiGRUNT~\cite{gu2022wigrunt} adopted attention-enhanced CNNs to automatically extract domain-robust features from CSI phase maps. PAC-CSI~\cite{su2023real} and other attention-based models~\cite{gu2023attention} achieved strong cross-domain performance through adaptive feature weighting. WiSGP~\cite{liu2023generalizing} improved robustness via data augmentation, while WiOpen~\cite{zhang2025wiopen} enhanced discriminability by exploiting geometric relationships within the feature space. EI~\cite{jiang2018towards} employed adversarial learning guided by physical domain labels to extract domain-invariant representations, and WiSR~\cite{liu2023wisr} extended this idea using subcarrier-based domain differentiation for improved generalization. i-Sample~\cite{zhou2024sample} enhanced adversarial generalization through data augmentation to achieve better stability. UniFi~\cite{liu2024unifi} proposed a unified framework that combines DFS and phase-based feature templates with a multi-view learning model to enhance cross-domain robustness. Its high-quality input generation strategy has proven remarkably effective, significantly advancing domain generalization research in WiFi sensing. By leveraging complementary perspectives from multiple transmitter–receiver antenna pairs, UniFi achieves strong cross-domain recognition. However, it requires at least two antenna pairs to function effectively, limiting its applicability in single-link or sparse-antenna settings. Other studies, such as Metaformer~\cite{sheng2024metaformer}, Wi-SFDAGR~\cite{yan2025wisfdar}, and Wi-Learner~\cite{feng2022wi}, achieved cross-domain recognition through domain alignment using limited target-domain data. While effective, these approaches depend on prior data collection in the target domain, which restricts scalability and practical deployment.
Compared with the current state-of-the-art UniFi~\cite{liu2024unifi}, which emphasizes high-quality input template design and multi-view feature fusion, \name approaches domain generalization from a distinct perspective. Its key contribution lies in revealing that directly partitioning domains based on physical configurations and applying a conventional adversarial framework can be suboptimal, or even detrimental, in WiFi sensing. \name systematically analyzes the underlying causes of this limitation and introduces a latent domain mining mechanism that enables adversarial generalization to achieve state-of-the-art performance. Although UniFi and \name tackle domain generalization through different means, they are inherently complementary and can be effectively integrated. For instance, \name also benefits from the physically grounded feature templates proposed in UniFi.

\section{Conclusion}

This paper systematically reexamined the limitations of conventional domain adversarial generalization methods in WiFi-based gesture recognition. Through theoretical analysis and empirical studies, we revealed that physical domain partitioning induces classification conflicts and manifold distortions, fundamentally constraining adversarial generalization effectiveness. To address these challenges, we proposed \name, a simple yet powerful framework that leverages WiFi latent domain mining to achieve more meaningful domain alignment. Extensive evaluations across public datasets and real-world scenarios demonstrate that \name consistently outperforms SOTA approaches in cross-domain generalization tasks. Moving forward, we envision integrating physical modeling and contextual information to further enhance latent domain discovery and mitigate complex manifold distortions in WiFi sensing.

\begin{acks}
This work is partially supported by the National Natural Science Foundation of China (No. 62502482, No. 62462015, No. 62422213), the Guizhou Provincial Basic Research Program(Natural Science) (Qiankehejichu-Youth[2024]345), the Beijing Nova Program (20240484641) and the Shandong Key Laboratory of Advanced Computing.
\end{acks}

\bibliographystyle{ACM-Reference-Format}
\bibliography{sample-base}

@String{Computing = "Computing" }

@String{Computer = "{IEEE} Computer" }

@String{Springer = "Springer-Verlag" }

@ARTICLE{zhang2023wital,
  author={Zhang, Xiang and Gu, Yu and Yan, Huan and Wang, Yantong and Dong, Mianxiong and Ota, Kaoru and Ren, Fuji and Ji, Yusheng},
  journal={IEEE Transactions on Human-Machine Systems}, 
  title={Wital: A COTS WiFi Devices Based Vital Signs Monitoring System Using NLOS Sensing Model}, 
  year={2023},
  volume={53},
  number={3},
  pages={629-641},
  publisher={IEEE}
  }

@article{huang2024keystrokesniffer,
  title={KeystrokeSniffer: An Off-the-Shelf Smartphone Can Eavesdrop on Your Privacy from Anywhere},
  author={Huang, Jinyang and Bai, Jia-Xuan and Zhang, Xiang and Liu, Zhi and Feng, Yuanhao and Liu, Jianchun and Sun, Xiao and Dong, Mianxiong and Li, Meng},
  journal={IEEE Transactions on Information Forensics and Security},
  year={2024},
  publisher={IEEE}
}

@article{wang2024xrf55,
  title={XRF55: A Radio Frequency Dataset for Human Indoor Action Analysis},
  author={Wang, Fei and Lv, Yizhe and Zhu, Mengdie and Ding, Han and Han, Jinsong},
  journal={Proceedings of the ACM on Interactive, Mobile, Wearable and Ubiquitous Technologies},
  volume={8},
  number={1},
  pages={1--34},
  year={2024},
  publisher={ACM New York, NY, USA}
}

@article{zhang2023imgfi,
  title={Imgfi: A high accuracy and lightweight human activity recognition framework using csi image},
  author={Zhang, Changsheng and Jiao, Wanguo},
  journal={IEEE Sensors Journal},
  volume={23},
  number={18},
  pages={21966--21977},
  year={2023},
  publisher={IEEE}
}

@inproceedings{li2021two,
  title={Two-stream convolution augmented transformer for human activity recognition},
  author={Li, Bing and Cui, Wei and Wang, Wei and Zhang, Le and Chen, Zhenghua and Wu, Min},
  booktitle={Proceedings of the AAAI conference on artificial intelligence},
  volume={35},
  number={1},
  pages={286--293},
  year={2021}
}

@article{feng2022wi,
  title={Wi-learner: Towards one-shot learning for cross-domain wi-fi based gesture recognition},
  author={Feng, Chao and Wang, Nan and Jiang, Yicheng and Zheng, Xia and Li, Kang and Wang, Zheng and Chen, Xiaojiang},
  journal={Proceedings of the ACM on Interactive, Mobile, Wearable and Ubiquitous Technologies},
  volume={6},
  number={3},
  pages={1--27},
  year={2022},
  publisher={ACM New York, NY, USA}
}

@inproceedings{sun2015widraw,
  title={Widraw: Enabling hands-free drawing in the air on commodity wifi devices},
  author={Sun, Li and Sen, Souvik and Koutsonikolas, Dimitrios and Kim, Kyu-Han},
  booktitle={Proceedings of the 21st Annual International Conference on Mobile Computing and Networking},
  pages={77--89},
  year={2015}
}

@inproceedings{venkatnarayan2018multi,
  title={Multi-user gesture recognition using WiFi},
  author={Venkatnarayan, Raghav H and Page, Griffin and Shahzad, Muhammad},
  booktitle={Proceedings of the 16th Annual International Conference on Mobile Systems, Applications, and Services},
  pages={401--413},
  year={2018}
}

@inproceedings{abdelnasser2015wigest,
  title={Wigest: A ubiquitous wifi-based gesture recognition system},
  author={Abdelnasser, Heba and Youssef, Moustafa and Harras, Khaled A},
  booktitle={2015 IEEE conference on computer communications (INFOCOM)},
  pages={1472--1480},
  year={2015},
  organization={IEEE}
}

@article{yu2018qgesture,
  title={QGesture: Quantifying gesture distance and direction with WiFi signals},
  author={Yu, Nan and Wang, Wei and Liu, Alex X and Kong, Lingtao},
  journal={Proceedings of the ACM on Interactive, Mobile, Wearable and Ubiquitous Technologies},
  volume={2},
  number={1},
  pages={1--23},
  year={2018},
  publisher={ACM New York, NY, USA}
}

@inproceedings{li2016wifinger,
  title={WiFinger: talk to your smart devices with finger-grained gesture},
  author={Li, Hong and Yang, Wei and Wang, Jianxin and Xu, Yang and Huang, Liusheng},
  booktitle={Proceedings of the 2016 ACM International Joint Conference on Pervasive and Ubiquitous Computing},
  pages={250--261},
  year={2016}
}

@inproceedings{ali2015keystroke,
  title={Keystroke recognition using wifi signals},
  author={Ali, Kamran and Liu, Alex X and Wang, Wei and Shahzad, Muhammad},
  booktitle={Proceedings of the 21st annual international conference on mobile computing and networking},
  pages={90--102},
  year={2015}
}

@article{zhang2020wisign,
  title={WiSign: Ubiquitous American Sign Language Recognition Using Commercial Wi-Fi Devices},
  author={Zhang, Lei and Zhang, Yixiang and Zheng, Xiaolong},
  journal={ACM Transactions on Intelligent Systems and Technology (TIST)},
  volume={11},
  number={3},
  pages={1--24},
  year={2020},
  publisher={ACM New York, NY, USA}
}

@article{gu2023attention,
  title={Attention-Based Gesture Recognition Using Commodity WiFi Devices},
  author={Gu, Yu and Yan, Huan and Zhang, Xiang and Wang, Yantong and Huang, Jinyang and Ji, Yusheng and Ren, Fuji},
  journal={IEEE Sensors Journal},
  volume={23},
  number={9},
  pages={9685--9696},
  year={2023},
  publisher={IEEE}
}

@inproceedings{li2024uwb,
  title={UWB-Fi: Pushing Wi-Fi towards Ultra-wideband for Fine-Granularity Sensing},
  author={Li, Xin and Wang, Hongbo and Chen, Zhe and Jiang, Zhiping and Luo, Jun},
  booktitle={Proceedings of the 22nd Annual International Conference on Mobile Systems, Applications and Services},
  pages={42--55},
  year={2024}
}

@ARTICLE{huang2021phaseanti,
  author={Huang, Jinyang and Liu, Bin and Miao, Chenglin and Lu, Yan and Zheng, Qijia and Wu, Yu and Liu, Jiancun and Su, Lu and Chen, Chang Wen},
  journal={IEEE Transactions on Mobile Computing}, 
  title={PhaseAnti: An Anti-Interference WiFi-Based Activity Recognition System Using Interference-Independent Phase Component}, 
  year={2023},
  volume={22},
  number={5},
  pages={2938-2954},
  publisher={IEEE}}

@article{su2023real,
  title={A Real-Time Cross-Domain Wi-Fi-based Gesture Recognition System For Digital Twins},
  author={Su, Jian and Mao, Qiankun and Liao, Zhenlong and Sheng, Zhengguo and Huang, Chenxi and Zhang, Xuedong},
  journal={IEEE Journal on Selected Areas in Communications},
  year={2023},
  publisher={IEEE}
}

@inproceedings{kim2021selfreg,
  title={Selfreg: Self-supervised contrastive regularization for domain generalization},
  author={Kim, Daehee and Yoo, Youngjun and Park, Seunghyun and Kim, Jinkyu and Lee, Jaekoo},
  booktitle={Proceedings of the IEEE/CVF International Conference on Computer Vision},
  pages={9619--9628},
  year={2021}
}

@article{wang2019joint,
  title={Joint activity recognition and indoor localization with WiFi fingerprints},
  author={Wang, Fei and Feng, Jianwei and Zhao, Yinliang and Zhang, Xiaobin and Zhang, Shiyuan and Han, Jinsong},
  journal={IEEE Access},
  volume={7},
  pages={80058--80068},
  year={2019},
  publisher={IEEE}
}

@article{liu2023generalizing,
  title={Generalizing Wireless Cross-Multiple-Factor Gesture Recognition to Unseen Domains},
  author={Liu, Shijia and Chen, Zhenghua and Wu, Min and Wang, Hao and Xing, Bin and Chen, Liangyin},
  journal={IEEE Transactions on Mobile Computing},
  year={2023},
  publisher={IEEE}
}

@article{liu2023wisr,
  title={WiSR: Wireless Domain Generalization Based on Style Randomization},
  author={Liu, Shijia and Chen, Zhenghua and Wu, Min and Liu, Chang and Chen, Liangyin},
  journal={IEEE Transactions on Mobile Computing},
  year={2023},
  publisher={IEEE}
}

@article{zeng2019farsense,
  title={FarSense: Pushing the range limit of WiFi-based respiration sensing with CSI ratio of two antennas},
  author={Zeng, Youwei and Wu, Dan and Xiong, Jie and Yi, Enze and Gao, Ruiyang and Zhang, Daqing},
  journal={Proceedings of the ACM on Interactive, Mobile, Wearable and Ubiquitous Technologies},
  volume={3},
  number={3},
  pages={1--26},
  year={2019},
  publisher={ACM New York, NY, USA}
}

@article{wu2020fingerdraw,
  title={Fingerdraw: Sub-wavelength level finger motion tracking with WiFi signals},
  author={Wu, Dan and Gao, Ruiyang and Zeng, Youwei and Liu, Jinyi and Wang, Leye and Gu, Tao and Zhang, Daqing},
  journal={Proceedings of the ACM on Interactive, Mobile, Wearable and Ubiquitous Technologies},
  volume={4},
  number={1},
  pages={1--27},
  year={2020},
  publisher={ACM New York, NY, USA}
}

@article{li2020wihf,
  title={WiHF: Gesture and user recognition with WiFi},
  author={Li, Chenning and Liu, Manni and Cao, Zhichao},
  journal={IEEE Transactions on Mobile Computing},
  volume={21},
  number={2},
  pages={757--768},
  year={2020},
  publisher={IEEE}
}

@inproceedings{niu2018fresnel,
  title={A fresnel diffraction model based human respiration detection system using COTS Wi-Fi devices},
  author={Niu, Kai and Zhang, Fusang and Chang, Zhaoxin and Zhang, Daqing},
  booktitle={Proceedings of the 2018 ACM international joint conference and 2018 international symposium on pervasive and ubiquitous computing and wearable computers},
  pages={416--419},
  year={2018}
}

@article{xu2024hypertracking,
  title={HyperTracking: Exploring the hyperbolic model for non-line-of-sight device-free Wi-Fi tracking},
  author={Xu, Xiaoqiang and Meng, Xuanqi and Tong, Xinyu and Liu, Xiulong and Xie, Xin and Qu, Wenyu},
  journal={Proceedings of the ACM on Interactive, Mobile, Wearable and Ubiquitous Technologies},
  volume={7},
  number={4},
  pages={1--26},
  year={2024},
  publisher={ACM New York, NY, USA}
}

@inproceedings{meng2023secur,
  title={Secur-Fi: A secure wireless sensing system based on commercial Wi-Fi devices},
  author={Meng, Xuanqi and Zhou, Jiarun and Liu, Xiulong and Tong, Xinyu and Qu, Wenyu and Wang, Jianrong},
  booktitle={IEEE INFOCOM 2023-IEEE Conference on Computer Communications},
  pages={1--10},
  year={2023},
  organization={IEEE}
}

@article{tong2024nne,
  title={NNE-tracking: A neural network enhanced framework for device-free Wi-Fi tracking},
  author={Tong, Xinyu and Ge, Weiping and Tian, Yichen and Liu, Zijuan and Liu, Xiulong and Qu, Wenyu},
  journal={IEEE Transactions on Mobile Computing},
  volume={23},
  number={9},
  pages={8981--8998},
  year={2024},
  publisher={IEEE}
}

@inproceedings{zhang2025diffloc,
author={Zhang, Xiang and Zhang, Jie and Yan, Huan and Huang, Jinyang and Ma, Zehua and Liu, Bin and Li, Meng and Chen, Kejiang and Guo, Qing and Zhang, Tianwei and Liu, Zhi},
booktitle = {The 34th USENIX Security Symposium},
title = {DiffLoc: WiFi Hidden Camera Localization Based on Electromagnetic Diffraction},
year = {2025},
address = {Seattle, WA, USA},
month =May}

@article{niu2021understanding,
  title={Understanding WiFi signal frequency features for position-independent gesture sensing},
  author={Niu, Kai and Zhang, Fusang and Wang, Xuanzhi and Lv, Qin and Luo, Haitong and Zhang, Daqing},
  journal={IEEE Transactions on Mobile Computing},
  volume={21},
  number={11},
  pages={4156--4171},
  year={2021},
  publisher={IEEE}
}

@inproceedings{jiang2018towards,
  title={Towards environment independent device free human activity recognition},
  author={Jiang, Wenjun and Miao, Chenglin and Ma, Fenglong and Yao, Shuochao and Wang, Yaqing and Yuan, Ye and Xue, Hongfei and Song, Chen and Ma, Xin and Koutsonikolas, Dimitrios and others},
  booktitle={Proceedings of the 24th annual international conference on mobile computing and networking},
  pages={289--304},
  year={2018}
}

@article{ben2010theory,
  title={A theory of learning from different domains},
  author={Ben-David, Shai and Blitzer, John and Crammer, Koby and Kulesza, Alex and Pereira, Fernando and Vaughan, Jennifer Wortman},
  journal={Machine learning},
  volume={79},
  pages={151--175},
  year={2010},
  publisher={Springer}
}

@article{sicilia2023domain,
  title={Domain adversarial neural networks for domain generalization: When it works and how to improve},
  author={Sicilia, Anthony and Zhao, Xingchen and Hwang, Seong Jae},
  journal={Machine Learning},
  volume={112},
  number={7},
  pages={2685--2721},
  year={2023},
  publisher={Springer}
}

@inproceedings{ganin2015unsupervised,
  title={Unsupervised domain adaptation by backpropagation},
  author={Ganin, Yaroslav and Lempitsky, Victor},
  booktitle={International conference on machine learning},
  pages={1180--1189},
  year={2015},
  organization={PMLR}
}

@article{gao2023wicgesture,
  title={WiCGesture: Meta-Motion Based Continuous Gesture Recognition With Wi-Fi},
  author={Gao, Ruiyang and Li, Wenwei and Liu, Jinyi and Dai, Shuyu and Zhang, Mi and Wang, Leye and Zhang, Daqing},
  journal={IEEE Internet of Things Journal},
  year={2023},
  publisher={IEEE}
}

@article{zhang2023unsupervised,
  title={Unsupervised domain adaptation for rf-based gesture recognition},
  author={Zhang, Bin-Bin and Zhang, Dongheng and Li, Yadong and Hu, Yang and Chen, Yan},
  journal={IEEE Internet of Things Journal},
  volume={10},
  number={23},
  pages={21026--21038},
  year={2023},
  publisher={IEEE}
}

@article{sheng2024cdfi,
  title={CDFi: Cross-Domain Action Recognition using WiFi Signals},
  author={Sheng, Biyun and Han, Rui and Cai, Hui and Xiao, Fu and Gui, Linqing and Guo, Zhengxin},
  journal={IEEE Transactions on Mobile Computing},
  year={2024},
  publisher={IEEE}
}

@article{chen2024wignn,
  title={WiGNN: WiFi-Based Cross-Domain Gesture Recognition Inspired by Dynamic Topology Structure},
  author={Chen, Yinan and Huang, Xiaoxia},
  journal={IEEE Wireless Communications},
  year={2024},
  publisher={IEEE}
}

@article{zhou2022domain,
  title={Domain generalization: A survey},
  author={Zhou, Kaiyang and Liu, Ziwei and Qiao, Yu and Xiang, Tao and Loy, Chen Change},
  journal={IEEE Transactions on Pattern Analysis and Machine Intelligence},
  volume={45},
  number={4},
  pages={4396--4415},
  year={2022},
  publisher={IEEE}
}

@article{gao2021towards,
  title={Towards position-independent sensing for gesture recognition with Wi-Fi},
  author={Gao, Ruiyang and Zhang, Mi and Zhang, Jie and Li, Yang and Yi, Enze and Wu, Dan and Wang, Leye and Zhang, Daqing},
  journal={Proceedings of the ACM on Interactive, Mobile, Wearable and Ubiquitous Technologies},
  volume={5},
  number={2},
  pages={1--28},
  year={2021},
  publisher={ACM New York, NY, USA}
}

@article{lu2024diversify,
  title={Diversify: A General Framework for Time Series Out-of-distribution Detection and Generalization},
  author={Lu, Wang and Wang, Jindong and Sun, Xinwei and Chen, Yiqiang and Ji, Xiangyang and Yang, Qiang and Xie, Xing},
  journal={IEEE Transactions on Pattern Analysis and Machine Intelligence},
  year={2024},
  publisher={IEEE}
}

@article{li2025wilife,
  title={WiLife: Long-Term Daily Status Monitoring and Habit Mining of the Elderly Leveraging Ubiquitous Wi-Fi Signals},
  author={Li, Shengjie and Liu, Zhaopeng and Lv, Qin and Zou, Yanyan and Zhang, Yue and Zhang, Daqing},
  journal={ACM Transactions on Computing for Healthcare},
  volume={6},
  number={1},
  pages={1--29},
  year={2025},
  publisher={ACM New York, NY}
}

@article{yi2025multi,
  title={Multi-Person Respiration Monitoring Leveraging Commodity Wi-Fi Devices},
  author={Yi, En-Ze and Niu, Kai and Zhang, Fu-Sang and Gao, Rui-Yang and Luo, Jun and Zhang, Da-Qing},
  journal={Journal of Computer Science and Technology},
  volume={40},
  number={1},
  pages={229--251},
  year={2025},
  publisher={Springer}
}

@inproceedings{zhang2025camlopa,
  title={CAMLOPA: A Hidden Wireless Camera Localization Framework via Signal Propagation Path Analysis},
  author={Zhang, Xiang and Zhang, Jie and Ma, Zehua and Huang, Jinyang and Li, Meng and Yan, Huan and Zhao, Peng and Zhang, Zijian and Liu,Bin and Guo, Qing and Zhang, Tianwei and Yu, NengHai},
  booktitle={2025 IEEE symposium on security and privacy (SP)},
  year={2025},
  organization={IEEE}
}

@inproceedings{chi2024rf,
  title={RF-diffusion: Radio signal generation via time-frequency diffusion},
  author={Chi, Guoxuan and Yang, Zheng and Wu, Chenshu and Xu, Jingao and Gao, Yuchong and Liu, Yunhao and Han, Tony Xiao},
  booktitle={Proceedings of the 30th Annual International Conference on Mobile Computing and Networking},
  pages={77--92},
  year={2024}
}

@article{hou2024rfboost,
  title={Rfboost: Understanding and boosting deep wifi sensing via physical data augmentation},
  author={Hou, Weiying and Wu, Chenshu},
  journal={Proceedings of the ACM on Interactive, Mobile, Wearable and Ubiquitous Technologies},
  volume={8},
  number={2},
  pages={1--26},
  year={2024},
  publisher={ACM New York, NY, USA}
}

@article{wang2025vr,
  title={VR-Fi: Positioning and Recognizing Hand Gestures via VR-embedded Wi-Fi Sensing},
  author={Wang, Hongbo and Li, Xin and Li, Jiachun and Zhu, Haojin and Luo, Jun},
  journal={IEEE Transactions on Mobile Computing},
  year={2025},
  publisher={IEEE}
}

@article{yi2024enabling,
  title={Enabling WiFi sensing on new-generation WiFi cards},
  author={Yi, Enze and Zhang, Fusang and Xiong, Jie and Niu, Kai and Yao, Zhiyun and Zhang, Daqing},
  journal={Proceedings of the ACM on Interactive, Mobile, Wearable and Ubiquitous Technologies},
  volume={7},
  number={4},
  pages={1--26},
  year={2024},
  publisher={ACM New York, NY, USA}
}

@article{zhou2024sample,
  title={i-Sample: Augment domain adversarial adaptation models for WiFi-Based HAR},
  author={Zhou, Zhipeng and Wang, Feng and Gong, Wei},
  journal={ACM Transactions on Sensor Networks},
  volume={20},
  number={2},
  pages={1--20},
  year={2024},
  publisher={ACM New York, NY}
}

@article{hassan2024adversarial,
  title={Adversarial AI applied to cross-user inter-domain and intra-domain adaptation in human activity recognition using wireless signals},
  author={Hassan, Muhammad and Kelsey, Tom and Rahman, Fahrurrozi},
  journal={Plos one},
  volume={19},
  number={4},
  pages={e0298888},
  year={2024},
  publisher={Public Library of Science San Francisco, CA USA}
}

@ARTICLE{yan2025wisfdar,
  author={Yan, Huan and Zhang, Xiang and Huang, Jinyang and Feng, Yuanhao and Li, Meng and Wang, Anzhi and Ou, Weihua and Wang, Hongbing and Liu, Zhi},
  journal={IEEE Internet of Things Journal}, 
  title={Wi-SFDAGR: WiFi-Based Cross-Domain Gesture Recognition via Source-Free Domain Adaptation}, 
  year={2025},
  volume={},
  number={},
  pages={1-1}}

@ARTICLE{cao2025widual,
  author={Cao, Chenhong and Ding, Yue and Dai, Miaoling and Gong, Wei and Zhao, Xibin},
  journal={IEEE Transactions on Mobile Computing}, 
  title={Real-time Cross-Domain Gesture and User Identification via COTS WiFi}, 
  year={2025},
  volume={},
  number={},
  pages={1-13}}

@ARTICLE{zhang2025wiopen,
  author={Zhang, Xiang and Huang, Jinyang and Yan, Huan and Feng, Yuanhao and Zhao, Peng and Zhuang, Guohang and Liu, Zhi and Liu, Bin},
  journal={IEEE Transactions on Human-Machine Systems}, 
  title={WiOpen: A Robust Wi-Fi-Based Open-Set Gesture Recognition Framework}, 
  year={2025},
  volume={55},
  number={2},
  pages={234-245}}

@article{sheng2024metaformer,
  title={MetaFormer: Domain-Adaptive WiFi Sensing with Only One Labelled Target Sample},
  author={Sheng, Biyun and Han, Rui and Xiao, Fu and Guo, Zhengxin and Gui, Linqing},
  journal={Proceedings of the ACM on Interactive, Mobile, Wearable and Ubiquitous Technologies},
  volume={8},
  number={1},
  pages={1--27},
  year={2024},
  publisher={ACM New York, NY, USA}
}

@article{liu2024unifi,
  title={UniFi: A Unified Framework for Generalizable Gesture Recognition with Wi-Fi Signals Using Consistency-guided Multi-View Networks},
  author={Liu, Yan and Yu, Anlan and Wang, Leye and Guo, Bin and Li, Yang and Yi, Enze and Zhang, Daqing},
  journal={Proceedings of the ACM on Interactive, Mobile, Wearable and Ubiquitous Technologies},
  volume={7},
  number={4},
  pages={1--29},
  year={2024},
  publisher={ACM New York, NY, USA}
}

@article{gu2022wigrunt,
  title={WiGRUNT: WiFi-enabled gesture recognition using dual-attention network},
  author={Gu, Yu and Zhang, Xiang and Wang, Yantong and Wang, Meng and Yan, Huan and Ji, Yusheng and Liu, Zhi and Li, Jianhua and Dong, Mianxiong},
  journal={IEEE Transactions on Human-Machine Systems},
  volume={52},
  number={4},
  pages={736--746},
  year={2022},
  publisher={IEEE}
}

@article{zhang2021widar3,
  title={Widar3. 0: Zero-effort cross-domain gesture recognition with Wi-Fi},
  author={Zhang, Yi and Zheng, Yue and Qian, Kun and Zhang, Guidong and Liu, Yunhao and Wu, Chenshu and Yang, Zheng},
  journal={IEEE Transactions on Pattern Analysis and Machine Intelligence},
  volume={44},
  number={11},
  pages={8671--8688},
  year={2021},
  publisher={IEEE}
}

\appendix
\section{Appendix}
\label{app}

For a distribution $P(X,Y)$ (denoted as $P$ in the following sections.) with an ideal binary labeling function $h$ and a hypothesis $f$. According to \cite{ben2010theory}, the error $\varepsilon (f)$ can be defined as:
\begin{equation}
    \varepsilon (f) = \mathbb{E}_{x\sim  P}|f(x)-h(x)|
\end{equation}

We also define the $\mathcal{H}$-divergence of two distribution $P$ and $Q$ over a space $\mathcal{X}$ and a hypothesis class $\mathcal{F}$ as follows \cite{ben2010theory}:
\begin{equation}
    d_\mathcal{H}(P,Q) = 2 \underset{f \sim \mathcal{F}}{sup}|Pr_{x \sim  P}(f(x)=1)-Pr_{x \sim  Q}(f(x)=1)| 
\end{equation}

We define the inter-domain class confusion rate $\eta$ as:
\begin{equation}
    \eta(\mathcal{S}^i, \mathcal{S}^j) = \mathbb{P}\left( y \neq y' \,\middle\vert\, x \sim \mathcal{S}^i,\, x' \sim \mathcal{S}^j,\, \| F(x) - F(x') \| \leq \delta \right)
\end{equation}
where $F(\cdot)$ is the data distribution description function, and $\delta$ is a small neighborhood radius.

\textbf{Proposition 2 (Classification Conflict):} If $\eta(\mathcal{S}^i, \mathcal{S}^j)$ is high, adversarial alignment in feature space increases $\sum_{k}^{}\varphi _k \varepsilon_{\mathcal{S}^k} (f)$ due to optimization conflict.

Suppose in region $\mathcal{R}$, adversarial training encourages $F(x) \approx F(x')$ for $x \sim \mathcal{S}_i$, $x' \sim \mathcal{S}_j$. If $y \neq y'$, then the task classifier $h$ is penalized for distinguishing them. This adds an irreducible error term proportional to $\eta(\mathcal{S}_i, \mathcal{S}_j)$:
\begin{equation}
    \sum_{k}^{}\varphi _k \varepsilon^*_{\mathcal{S}^k} (f) \geq \sum_{k}^{}\varphi _k \varepsilon_{\mathcal{S}^k} (f) + C\cdot\eta(\mathcal{S}_i, \mathcal{S}_j)
\end{equation}
where $C$ is a constant depending on the classification margin.

\textbf{Proposition 3 (Manifold Distortion):} If two physical domains $\mathcal{S}^i, \mathcal{S}^j$ are separated by a complex manifold trajectory, adversarial alignment directly between $\mathcal{S}^i$ and $\mathcal{S}^j$ introduces feature space distortion, causing $d_\mathcal{H}(\mathcal{S}_i, \mathcal{T})$ to be misestimated.

Let $\mathcal{M}$ denote the true manifold of WiFi sensing signals. Assume $\mathcal{S}^i$ and $\mathcal{S}^j$ lie on distant regions of $\mathcal{M}$. Adversarial alignment minimizes $|F(x) - F(x')|$ for $x \sim \mathcal{S}^i, x' \sim \mathcal{S}^j$. However, the correct manifold geodesic distance $d_\mathcal{M}(x, x')$ is hard to compute without the guidance of precise physical model. Forced alignment compresses this distribution distance, distorting the local geometry of $\mathcal{M}$ in real-world space. Thus, the estimation of $d_\mathcal{H}$ under $f(\cdot)$ no longer reflects true domain divergence, weakening the bound:
\begin{equation}
    \varepsilon_\mathcal{T} (f) \le \lambda +  \sum_{k}^{}\varphi _k \varepsilon_{\mathcal{S}^k} (f) + \frac{1}{2} d^{distorted}_\mathcal{H}(\mathcal{S},\mathcal{T}) + \frac{1}{2}\max_{i,j} d_\mathcal{H}(\mathcal{S}^i,\mathcal{S}^j)
\end{equation}

\end{document}